\documentclass[prb,preprint,superscriptaddress,floatfix]{revtex4-1}
\usepackage{epsfig}
\usepackage{float}
\usepackage{subfigure}
\usepackage{graphicx}
\usepackage{amssymb}
\usepackage{amsmath}

\usepackage{dcolumn}% Align table columns on decimal point
\usepackage{bm}% bold math
\usepackage{ulem}
\usepackage{epstopdf}
\usepackage{color}
\usepackage{soul}

\begin{document}

\title{Thermally activated switching at long time scales in exchange-coupled magnetic grains}

\author{Ahmad M. Almudallal}
\affiliation{Department of Physics and Physical Oceanography,
Memorial University of Newfoundland, St. John's, NL, A1B 3X7, Canada}

\author{J. I. Mercer}
\affiliation{Department of Computer Science, 
Memorial University of Newfoundland, St. John's, NL, A1B 3X7, Canada}

\author{J. P. Whitehead}
\affiliation{Department of Physics and Physical Oceanography,
Memorial University of Newfoundland, St. John's, NL, A1B 3X7, Canada}

\author{M. L. Plumer}
\affiliation{Department of Physics and Physical Oceanography,
Memorial University of Newfoundland, St. John's, NL, A1B 3X7, Canada}

\author{J. van Ek}
\affiliation{Western Digital Corporation,
Jan Jose, California 94588, USA}

\author{T. J. Fal}
\affiliation{Department of Physics and Energy Science,
University of Colorado Colorado Springs, CO, 80918, USA}

\date{\today}

\begin{abstract}

Rate coefficients of the Arrhenius-N\'{e}el form are calculated for thermally activated magnetic moment reversal for dual layer exchange-coupled composite (ECC) media based on the Langer formalism and are applied to study the sweep rate dependence of $MH$ hysteresis loops as a function of the exchange coupling $I$ between the layers. The individual grains are modelled as two exchange coupled Stoner-Wohlfarth particles  from which the minimum energy paths connecting the minimum energy states are calculated using a variant of the string method and the energy barriers and attempt frequencies calculated as a function of the applied field. The resultant rate equations describing the evolution of an ensemble of non-interacting ECC grains  are then integrated numerically in an applied field with constant sweep rate $R=-dH/dt$ and the magnetization calculated as a function of the applied field $H$. $MH$ hysteresis loops are presented for a range of values $I$ for sweep rates $10^{5}\:\mathrm{Oe/s} \le  R \le 10^{10}\:\mathrm{Oe/s}$ and a figure of merit (FOM) that quantifies the advantages of ECC media is proposed. $MH$ hysteresis loops are also calculated based on the stochastic Landau-Lifshitz-Gilbert equations for $10^{8}\:\mathrm{Oe/s} \le  R \le 10^{10}\:\mathrm{Oe/s}$ and are shown to be in good agreement with those obtained from the direct integration of rate equations. The results are also used to examine the accuracy of certain approximate models that reduce the complexity associated with the Langer based formalism and which provide some useful insight into the reversal process and its dependence on the coupling strength and sweep rate. Of particular interest is the clustering of minimum energy states that are separated by relatively low energy barriers into ``metastates". It is shown that while approximating the reversal process in terms of ``metastates" results in little loss of accuracy, it can reduce the run time of a Kinetic Monte Carlo (KMC) simulation of the magnetic decay of an ensemble of dual layer ECC media by $2\sim 3$ orders of magnitude. The essentially exact results presented in this work for two coupled grains are analogous to the Stoner-Wohlfarth model of a single grain and serve as an important precursor to KMC based simulation studies on systems of interacting dual layer ECC media. 

\end{abstract}

\maketitle

\section{Introduction}

After many decades, the Landau-Lifshitz-Gilbert (LLG) equation continues to provide the foundation for micromagnetic modeling of the dynamic evolution of granular magnetic material, with an increasing number of applications devoted to the study of the effects of thermal fluctuations. \cite{plumer1,brown} The LLG equation is commonly used to study granular recording media but it is limited to the study of phenomena over relatively short time scales ($ms$). Modern exchange-coupled composite (ECC) recording media is composed of a high anisotropy `hard' layer exchange coupled to one or more lower anisotropy `soft' layers. \cite{victora,kapoor,wang,suess,richter,choe,plumer2}   One of the most important applications of micromagnetic modeling is the characterization of recording media through $MH$ hysteresis loops.  Often, model parameters are determined by fitting results to experimental data. Inconveniently, experimental hysteresis loops typically require minutes to hours to complete, time scales that are outside the range of standard LLG simulations. In addition, the thermally activated decay of recorded bits requires a micromagnetic model that is valid over much longer time scales (years). Various scaling arguments, based on the Arrhenius-N\'{e}el law, have been proposed as a means to extrapolate LLG results to longer times which appear useful for older single layer type recording media.\cite{xue}   Thermally activated processes in ECC media, however, are more complex and the simple scaling arguments appear to break down in this case. \cite{plumer3}   

For the purpose of studying long-time scale micromagnetics governed by thermally activated processes, Kinetic Monte Carlo (KMC) methods have proven useful. \cite{chantrell,kanai,lu,charap,parker} In Ref.\;\onlinecite{fal1}, a KMC algorithm to study long-time scale thermally activated grain reversal of single layer recording media was described. The Arrhenius-N\'{e}el expression for the rate coefficients between the minimum energy states of the individual grains was used to calculate the time between successive reversals. The minimum energy states and the energy barriers separating them were calculated using a modified version of the Wood analytic  expression for single Stoner-Wohlfarth particles\cite{wood} (SWPs)  which includes the effective exchange and magnetostatic fields from neighbouring grains. For weakly interacting recording media, the effective field approximation appears valid. For the attempt frequency,  the temperature and field dependent formula of Wang and Bertram,\cite{bertram} based on a single energy barrier was used. This algorithm was subsequently used to study the magnetic $MH$ hysteresis loops of high anisotropy magnetic recording media at both short and long time scales over a wide range of temperatures relevant to heat assisted magnetic recording. \cite{plumer4}  Good agreement between the KMC results and those from LLG simulations at relatively short time scales was demonstrated.

In Ref.\;\onlinecite{fal2}, this KMC algorithm was applied to study $MH$ hysteresis loops of dual layer ECC recording media at finite temperature and long time scales in which the effect of ECC interlayer exchange coupling was treated in the same way as the intralayer exchange coupling by means of an effective field.  ECC media for practical applications has a relatively strong exchange coupling and it is not evident that treating the intralayer coupling through an effective field is a good approximation\cite{choe} for this purpose as it ignores the correlated nature of the rotation of the layers in the reversal process.  In addition, the expression used for the attempt frequency is based on a single energy barrier, is unlikely to be valid for multi-layer ECC media at relevant (moderate) coupling strengths. The absence of simple Arrhenius-N\'{e}el-type scaling between thermal and temporal effects for ECC media supports these conclusions.\cite{plumer3} 

In the present work, we study the reversal process for two interacting magnetized grains which treats the correlated reversal process in a more systematic way. The approach includes the complete set of minimum energy states for the ECC grains, while the calculation of the rate coefficients, based on the Langer formalism,\cite{langer} takes into account the complex minimum energy paths (MEPs) connecting them. The resultant energy barriers and attempt frequencies provide a comprehensive treatment of reversal process that is applicable to both weak and strong interlayer coupling. 

In order to study statistical effects, we also consider an ensemble of non-interacting exchange coupled dual layer grains. This has the benefit that we can describe the evolution of the ensemble from some initial distribution of states in terms of a system of rate equations that can be integrated numerically. In particular we present a series of $MH$ hysteresis loops calculated at constant sweep rate over a range of coupling constants to examine the effect of the exchange parameter and sweep rate on the $MH$ hysteresis loops. This work compliments and extends previous studies of dual-grain-reversal energy landscapes~\cite{dualE} and formulations of the dual-grain attempt frequency, \cite{bertram,f0}  and serves as an precursor to our formulation and application of the combined MEP-KMC algorithm to study interacting N$\times$N$\times$2 ECC thin films which includes both magnetostatic and intralayer exchange interactions. \cite{mepkmc} 
 
In the following section, we discuss the energy landscapes for a dual layer grain, which we model as a system consisting of two coupled Stoner-Wohlfarth particles (SWPs) in a magnetic field. We briefly outline how the rate coefficients may be calculated based on the Langer formalism for such a system of exchange coupled SWPs in the strong and weak coupling regimes. The details of the Langer formalism are presented in the Appendix. In Sec.~\ref{MHloops}, we show how the equations may be integrated and the rate coefficients determined from the MEP calculated using a variant of the so-called ``string method",\cite{henkelmana,weinan1,weinan2} and a series of $MH$ hysteresis loops are presented for various sweep rates and couplings. A figure of merit to assist in the evaluation of the benefits of ECC media is proposed based on the ratio of the switching field and the energy barriers and is calculated as a function of the exchange coupling for several different sweep rates. In Sec.~\ref{llg}, we compare the $MH$ hysteresis loops obtained from the rate equation approach with those obtained using stochastic LLG simulation over a range of sweep rates, where both approaches should be valid. The good agreement between the results from the two methods gives confidence that the results obtained from the rate equation approach, which can be extended to very low sweep rates, are essentially of the same quality as those obtained from stochastic LLG, which is restricted to very high sweep rates. 

In addition to the MEP based calculations of the thermally activated dual-grain reversal, we also examine the validity of two important approximation schemes. The first, based on approximations to the MEP allow us to obtain analytical expressions for the energy barrier and attempt frequency in the strong and weak coupling limits, respectively. A comparison of results obtained based on this scheme with MEP calculations are presented in Sec.~\ref{results} and show good agreement over a range of couplings and sweep rates. The second approximation method exploits the fact that in dual layer ECC media, the rate coefficients calculated from the MEPs separating pairs of energy minima can differ by orders of magnitude. A direct consequence of this is that pairs of minima will equilibrate on time scales that are significantly shorter than the time taken to complete a single sweep, or portion of a sweep. In such cases it is possible to combine the two minimum energy states into a single ``metastate". In Sec.~\ref{am-sec}, we show how this approach can be used to reduce a 4-state model to an equivalent 2-metastate representation that gives essentially the same results. While the difference in computation time required to solve the rate equations for the 4-state model and the equivalent 2-metastate model is negligible, the same is not true when we use KMC to calculate $MH$ hysteresis loops that include the interlayer interaction between the layers. In this case, the large variation in rate coefficients causes the KMC algorithm to slow to a crawl, giving rise to what we refer to as ``stagnation". In Sec.~\ref{metastatesKMC}, we illustrate the effects of stagnation by applying the KMC method to compute magnetization decay for an ensemble of non-interacting ECC grains.  It is shown that by combining certain pairs of states into metastates, the calculation speeds up by a factor of 600 with negligible loss in accuracy. KMC studies on dual layer ECC grains that include the magnetostatic and intralayer exchange interactions show that the clustering of minimum energy states separated by low energy barriers into metastates to avoid the effects of stagnation is critical to the successful application of KMC to systems of interest in magnetic recording media.  \cite{mepkmc} 

\section{Energy Landscapes and rate equations in ECC Media}
\label{energyLandscapes}

ECC media consists of magnetic grains with different layers of varying anisotropy strength and moderate exchange interactions between these layers.\cite{victora,kapoor,wang,suess,richter,choe}  The desired effect is to be able to use the very strong anisotropy of a hard layer to enhance the grains thermal stability and hence prevent data loss due to thermally activated grain reversal. The hard layer is then exchange coupled to a layer with lower anisotropy, the soft layer. The soft layer will respond more readily to a switching field and the exchange coupling interaction will make switching the hard layer easier. The result is a thermally stable grain that can be reoriented by using an applied field at lower magnitudes than would be required if just the hard layer were presented. 

In this work, we consider an ensemble of two exchange coupled grains. In order to focus on the role of the exchange coupling between the layers and to allow a semi-analytical treatment of the system, we neglect the lateral exchange interaction between the grains and the magnetostatic interaction. 

The grains are cubic with a side length $a$=6\:nm stacked along the $z$-axis.  The dimensions of the grains are such that each may be treated as a single domain ferromagnet and may be modelled as two exchange coupled SWPs which we label as $a$ and $b$. The energy of a single grain is therefore written in terms of the normalized magnetization vectors $\hat m_i = \vec M_i/M_i$,
\begin{align}
E=&-K_a v_a \left(\hat m_a\cdot\hat n_a\right)^2 -K_b v_b \left(\hat m_b\cdot\hat n_b\right)^2
-IA (\hat m_a \cdot \hat m_b) \nonumber\\
&- \mu_0 \vec H\cdot \left(M_a v_a \hat m_a +   M_b v_b \hat m_b\right),
\end{align}
where $\vec H$ denotes the applied field and $K_i$, $v_i$, $\hat n_i$ and $M_i$ denote the anisotropy constant, volume, anisotropy axis and the saturation magnetization of the $i^\text{th}$ grain, respectively. The grains we consider are comprised of a soft layer with $M_a$=4.0$\times 10^5$\:A/m and $K_a$=1.5$\times 10^5$\:J/m$^3$, and a hard layer with $M_b$=5.4$\times 10^5$\:A/m and $K_b$=3.0$\times 10^5$\:J/m$^3$. The layers are coupled through ferromagnetic exchange expressed in terms of the coupling constant $I$ and interfacial area $A=a^2$. Assuming that both the anisotropy axis $\hat n_i$ and the field $\vec H$ are aligned perpendicular to the plane, then the SWP energy for a single grain may be written in spherical coordinates as,\begin{align}
E=&-K_av_a \sin^2\theta_a  -K_b v_b \sin^2\theta_b
- \mu_0 H \left(M_a v_a \sin\theta_a  +   M_b v_b \sin\theta_b \right)\nonumber\\
&-IA \left( \sin\theta_a\sin\theta_b + \cos\left(\phi _a -\phi_b\right) \cos\theta_a \cos\theta_b\right),
\end{align}
where $\theta_a$ and $\theta_b$ denote the polar angles measured relative to the $xy$ plane and $\phi_a$ and $\phi_b$ denote the azimuthal angles associated with the grains $a$ and $b$, respectively. 

To understand how the energy of a grain depends on the variables $\{\theta_a,\phi_a,\theta_b,\phi_b\}$, we note first that it is invariant under rotation about the $z$-axis and thus depends only on the three independent variables $\{\theta_a,\theta_b,\phi_b-\phi_a\}$. Also since the exchange coupling between the layers is positive, the energy is minimized when $\phi_a = \phi_b$, we therefore find it useful to plot the two dimensional subspace defined by $\phi_a = \phi_b$, which refer to as the ``minimum energy surface". We consider four specific cases in some detail corresponding  to $I=2.0 \times 10^{-3}\:\mathrm{J/m^2}$ and $I=0.5 \times 10^{-3}\:\mathrm{J/m^2}$ for both $H=0$ and $\mu_0 H=4\:\mathrm{kOe}$.

\subsection{Strong Exchange Coupling}
\label{strongExchange}

\begin{figure}[ht]
\subfigure{
\centering\includegraphics[clip=true, trim=0 0 0 0, height=7.0cm]{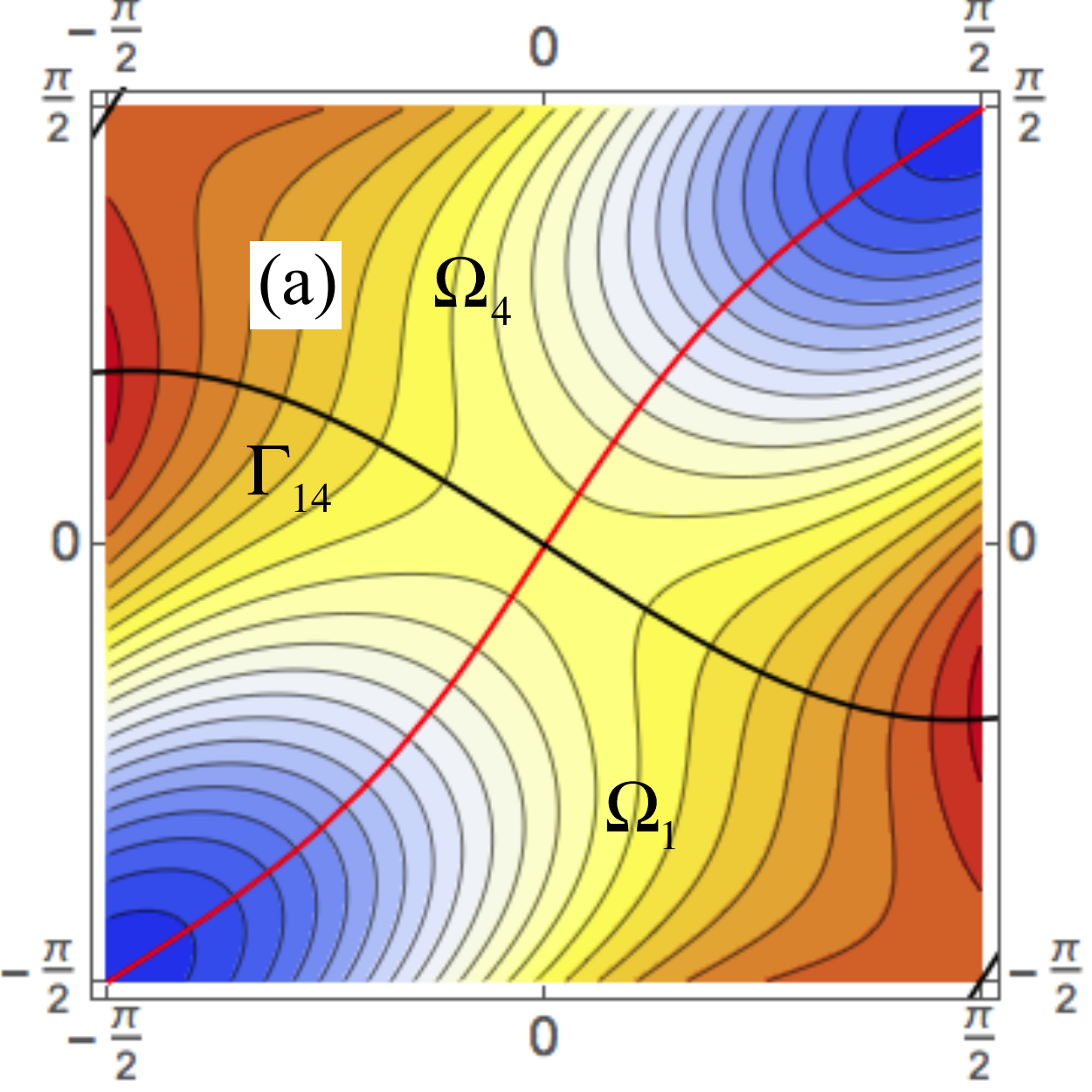}
\label{Almudallal_fig01_a}
}
\subfigure{
\centering\includegraphics[clip=true, trim=0 0 0 0, height=7.0cm]{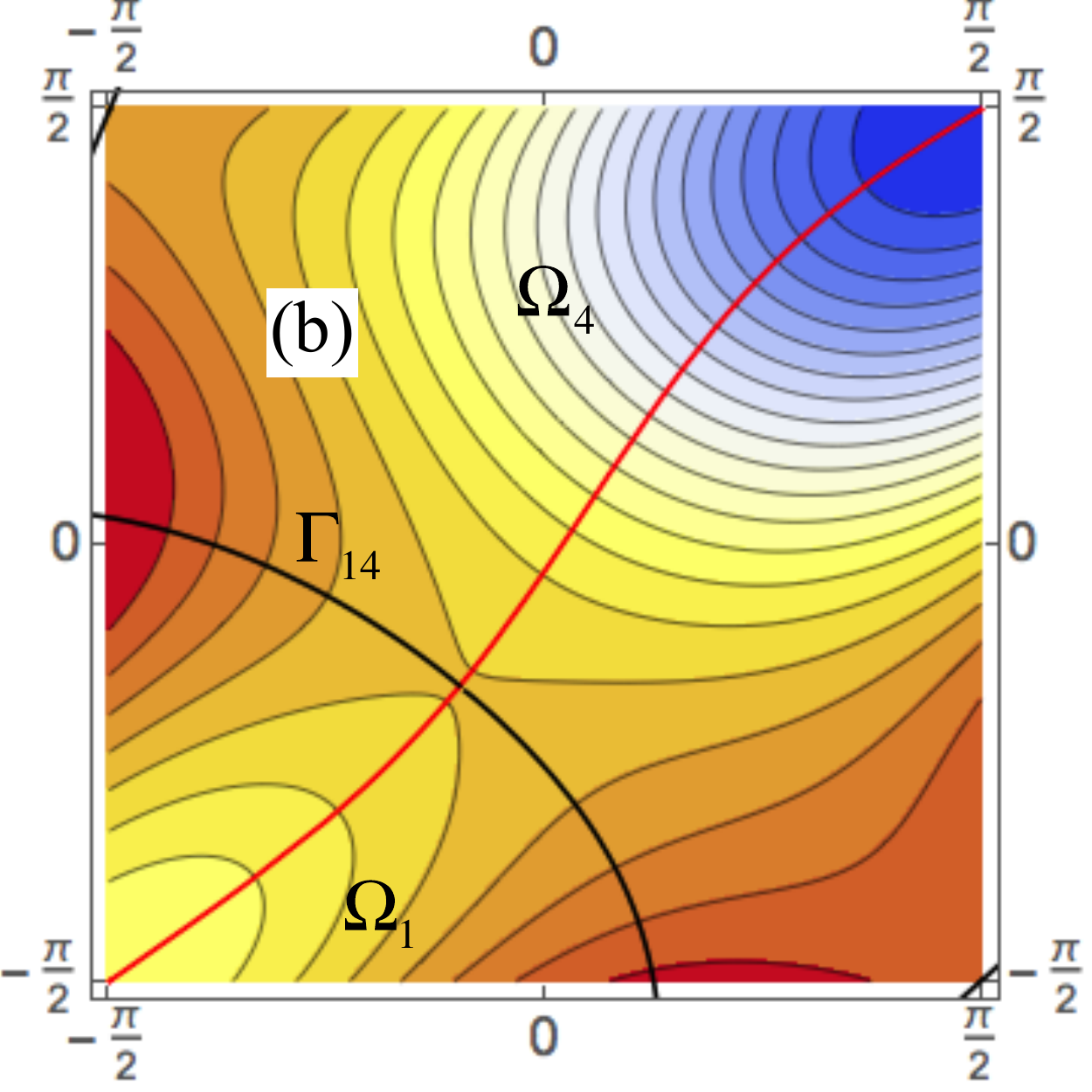}
\label{Almudallal_fig01_b}
}
\caption{The contour plot of the minimum energy surface for the strong coupling case $I$=2.0$\times 10^{-3}$\:J/m$^2$ for (a) $H$=0 and (b) $H$=4\:kOe. The black lines indicate the boundary separating the two basins of attraction $\Omega_1$ and $\Omega_4$ which we denote by $\Gamma_{14}$. The red lines indicate the MEP connecting the minimum energy states $\sigma_1$ and $\sigma_4$. The lines cross at the saddle point indicated by $s_{14}$.}
\label{Almudallal_fig01}
\end{figure}

In Fig.~\ref{Almudallal_fig01_a}, the contour plot of the minimum energy surface over the range $-\pi/2 < \theta_a < \pi/2$ and $-\pi/2 < \theta_b < \pi/2$ for $H=0$ and $I=2.0 \times 10^{-3}\:\mathrm{J/m^2}$ is presented. The energy landscape  shows two minima corresponding to two stable states with the magnetic spins aligned ferromagnetically along the $z$-axis. We refer to this as the strong exchange coupling regime and denote the two minimum energy states $\{\theta_a,\theta_b\} =\{-\pi/2,-\pi/2\} $ and $\{\pi/2,\pi/2\} $ as $\sigma_1$ and $\sigma_4$, respectively. We note that in the absence of a field, the energy landscape is symmetric under spin inversion and hence the minimum energy states are degenerate $E_1 = E_4$. 

The contour plot of the minimum energy surface of a grain is presented in Fig.~\ref{Almudallal_fig01_b} for $H=4\:\mathrm{kOe}$. The minimum energy landscape again shows two minima located at $\sigma_1$ and $\sigma_4$ however, because of the applied field, the energies are no longer degenerate and $E_4 < E_1$ so the system now has one stable minimum at $\sigma_4$ and a metastable minimum at $\sigma_1$.

Associated with each of the local minimum energy states $\sigma_\alpha$ is a basin of attraction defined as the region of phase space comprising the states that evolve asymptotically to the state $\sigma_\alpha$. We denote the basin of attraction associated with the state $\sigma_\alpha$ as $\Omega_\alpha$. Figure~\ref{Almudallal_fig01_a} and Fig.~\ref{Almudallal_fig01_b} show the boundary separating the two basins of attractions $\Omega_1$ and $\Omega_4$ which we denote by $\Gamma_{14}$. 

The probability distribution of the grains in phase space  is given by the probability density $\rho(x,t)$, where $x$ denotes a vector that specifies a point in phase space in terms of some generalized coordinates (e.g. $(x^1,x^2,x^3,x^4) = (\theta_a,\phi_a, \theta_b,\phi_b$)). The evolution of the probability density is given by the Fokker-Planck equation (FPE).  For the energy and time scales of interest to us, the system will be in local equilibrium. Local equilibrium assumes that, except for a narrow crossover region $\Delta_{14}$ that runs along the boundary $\Gamma_{14}$,  the probability density $\rho(x,t)$ is given by the Boltzmann distribution,
\begin{align}
\rho(x,t) \approx c_\alpha(t) \exp\left(-\frac{E(x,t)}{k_B T} \right) \, \text{for} \,x \in \Omega_\alpha - \Delta_{14},
\label{localEquilibrium}
\end{align}
where (except in the case of thermal equilibrium) $c_1 \ne c_2$. The probability $p_\alpha$ that a grain in the ensemble is located in $\Omega_\alpha$ is therefore given by,
\begin{align}
p_\alpha(t) = c_\alpha(t) \int_{\Omega_\alpha} \exp\left( -\frac{E(x,t)}{k_BT}\right) d\Omega\equiv c_\alpha(t)\mathcal{Z}_\alpha\label{localEq}.
\end{align}

In the crossover region $\Delta_{14}$, the system is not in equilibrium and the probability density is given by the more general form $\rho(x,t) = c(x,t) \exp(-E(x,t)/k_BT)$, where the crossover function $c(x,t)$ is obtained from the FPE and interpolates between the coefficients $c_{1}(t)$ and $c_{4}(t)$ defined by  Eq.~\eqref{localEquilibrium}. The inhomogeneous nature of $c(x,t)$ in the crossover region gives rise to a net flux of probability across the boundary  that is driven by the thermal fluctuations. For the energy scales of interest, this probability flux is concentrated at the point of minimum energy on the boundary $\Gamma_{14}$. This point, which we denote by $s_{14}$, is a saddle point with $\partial E/\partial x^\mu = 0$ and a Hessian matrix $|| \partial^2 E/\partial x^\mu \partial x^\nu||$ that has two positive eigenvalues, one negative eigenvalue and a zero eigenvalue (the latter arising as a consequence of the rotational symmetry of the energy about the axis perpendicular to the plane). 

The rate at which the particles in the ensemble make the transition from $\sigma_\alpha \to \sigma_\beta$ may be expressed in terms of the rate constant as, 
\begin{align}
\mathcal{I}_{\alpha\to\beta} = - r_{\alpha\beta} \,p_\alpha(t)\label{probCurrent},
\end{align}
where the rate constants $r_{\alpha\beta}$ are of the Arrhenius-N\'{e}el form,
\begin{align}
r_{\alpha\beta} = f_{\alpha\beta} \exp\left( -\frac{\Delta E_{\alpha\beta}}{k_BT}\right)\label{ArrheniusNeel},
\end{align}
where the energy barrier $\Delta E_{\alpha\beta} = E(s_{14} ) - E(\sigma_\alpha)$ and the attempt frequency, $f_{\alpha\beta}$, may be calculated from the crossover function $c_(x,t)$ in the neighbourhood of the saddle point $s_{\alpha\beta}$, and may be expressed as,
\begin{align}
f_{\alpha\beta}= \frac{\alpha_0}{1+\alpha_0^2}\sqrt{\frac{\tilde{g}(s)}{\bar g(\alpha)}} \frac{\gamma_B}m  G_U| \kappa|\sqrt{\frac{1}{2\pi k_BT}\frac{\eta_1\eta_2\eta_3\eta_4}{|\lambda_1\lambda_2\lambda_4|}},\label{attemptFrequency}
\end{align}
where {$\alpha_0$ is the damping constant, $\gamma_B$ is the gyromagnetic ratio, $m=M_a v_a+M_b v_b$, $\lambda_i$ and $\eta_i$ are the eigenvalues of the Hessian matrix $||\partial^2 E/\partial x^\mu \partial x^\nu||$ calculated at the saddle point, $s_{\alpha\beta}$, and the minimum energy state, $\sigma_\alpha$, respectively, $\kappa^{-1/2}$ characterizes the width of the crossover region $\Delta_{\alpha\beta}$ in the vicinity of the saddle point $s_{\alpha\beta}$, and the quantities $\tilde g(s)$, $\bar g(\alpha)$, and $G_U$ are related to metric associated with the particular coordinate system (or systems) used in the derivation. The details of the derivation of Eq.~\eqref{attemptFrequency} are presented in the Appendix.

The time dependence of the probabilities $p_\alpha(t)$ can be calculated from the rate equations,
\begin{align}
\label{rateEqs1}
\frac{dp_1}{dt} &= - r_{14} p_1 + r_{41}p_4 \\
\frac{dp_4}{dt} &= - r_{41} p_4+ r_{14}p_1.
\label{rateEqs2}
\end{align} 

\subsection{Weak Exchange Coupling}
\label{weakExchange}

Figures \ref{Almudallal_fig02_a} and \ref{Almudallal_fig02_b} show the corresponding energy landscapes for the case $I = 0.5\times 10^{-3}\:\mathrm{J/m^2}$ for $H=0$ and $4\:\mathrm{kOe}$. Both cases have four minimum energy configurations corresponding to the ferromagnetically aligned states $\{\theta_a,\theta_b\} = \{\mp \pi/2, \mp \pi/2\}$ and the antiferromgnetically aligned states $\{\theta_a,\theta_b\} =\{\pm \pi/2, \mp \pi/2\}$. We denote the minimum energy states $\{\mp \pi/2, \mp \pi/2\}$ by $\sigma_1$ and $\sigma_4$, as in the strong coupling case discussed above, and the antiferromagnetic states $\{\pi/2,-\pi/2\}$ and $\{-\pi/2,\pi/2\}$ as $\sigma_2$ and $\sigma_3$, respectively. We refer to this as the weak exchange coupling regime.  As before, in the absence of an applied field, the system is invariant under a spin inversion and we have  the following degeneracies $E_1=E_4$ and $E_2=E_3$. Figures~\ref{Almudallal_fig02_a} and \ref{Almudallal_fig02_b} also show the basin boundaries $\Gamma_{\alpha\beta}$ separating the basins of attraction $\Omega_\alpha$ associated with the energy minima $\sigma_\alpha$. 

\begin{figure}[ht]
\subfigure{
\centering\includegraphics[clip=true, trim=0 0 0 0, height=7.0cm]{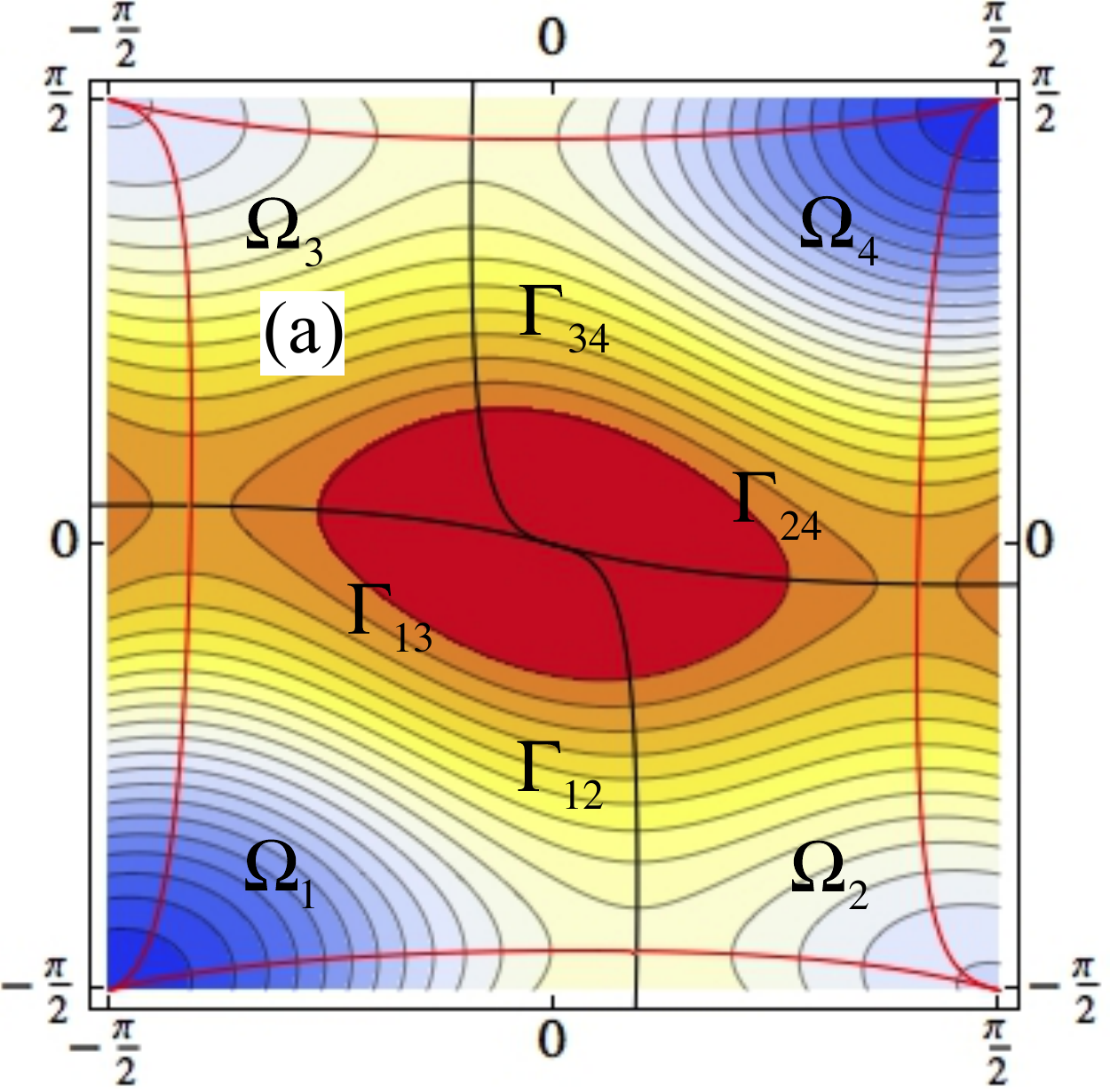} 
\label{Almudallal_fig02_a}
}
\subfigure{
\centering\includegraphics[clip=true, trim=0 0 0 0, height=7.0cm]{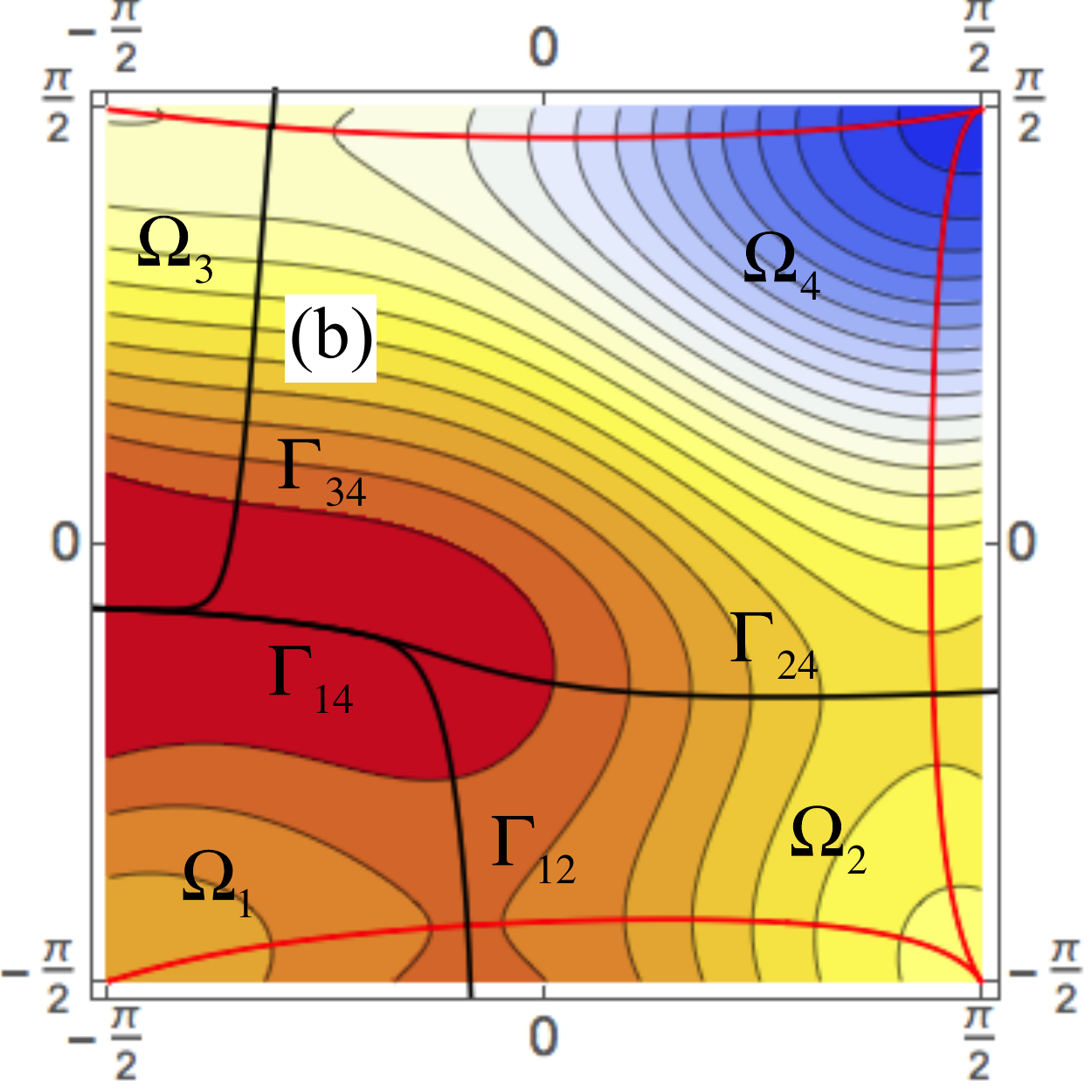}
\label{Almudallal_fig02_b}
}
\caption{The contour plot of the minimum energy surface for the weak coupling case $I=0.5\times 10^{-3}\:\mathrm{J/m^2}$ for (a) $H=0$ and (b) $H=4\:\mathrm{kOe}$. The black lines indicate the boundary separating each pair of basins of attractions $\{\Omega_\alpha,\Omega_\beta\}$ and is denoted by $\Gamma_{\alpha\beta}$. The red lines indicate the MEP connecting the four minima. The lines cross at the saddle points $s_{\alpha\beta}$.}
\label{Almudallal_fig02}
\end{figure}

For the energy and time scales of interest, the system will be in local equilibrium and we can therefore define the probabilities that a grain is located in $\Omega_\alpha$ as, 
\begin{align}
p_\alpha(t) = c_\alpha(t) \int_{\Omega_\alpha} \exp\left( -\frac{E(x,t)}{k_BT}\right) d\Omega \equiv c_\alpha(t) \mathcal{Z}_{\alpha}(t),
\end{align}
and the probability flux $\mathcal{I}_{\alpha\to\beta}$ between the basins of attractions $\{\Omega_\alpha,\Omega_\beta\}$ will be concentrated at the saddle point $s_{\alpha\beta}$ and may be expressed in terms of the rate constants $r_{\alpha\beta}$ of the forms given by Eq.~\eqref{probCurrent}. The formalism presented in the Appendix applies equally well to the systems with multiple energy minima.  The rate equations given by Eqs.~\eqref{rateEqs1} and~\eqref{rateEqs2} for strong coupling regime may then be written to include the case of multiple (i.e. more than two) minima as, 
\begin{align}
\frac{dp_\alpha}{dt} &= - \sum_{\beta = 1}^{N_s}\left( r_{\alpha\beta} p_\alpha - r_{\beta\alpha}p_\beta\right). 
\label{rateEq}
\end{align}
In applying the above formula we note that $r_{\alpha\alpha} \equiv 0$ and that the number of minimum energy states, $N_s$, will depend on the strength of the applied field, ranging from 1 to 2 in the strong coupling regime and from 1 to 4 in the weak coupling regime. 

\section{The Minimum Energy Path and the evaluation of $MH$ Hysteresis Loops} 
\label{MHloops}

To evaluate the $MH$ hysteresis loop for a layer of non-interacting ECC grains using the rate equations given by Eq.~\eqref{rateEq}, we consider that at some initial time, $t=t_i$, the system is fully saturated $p_1(t=t_i) = 1$ in a large positive applied field with only one minimum energy state $\sigma_1$. The field is then reduced at a constant rate $dH/dt = -R$ until the system is again fully saturated in the opposite direction at time $t_f$, $p_4(t=t_f)=1$. Since the rate of change of the applied field is constant, we have that $dp_\alpha/dt = - Rdp_\alpha/dH$ and the rate equations may be written as, 
\begin{align}
\frac{dp_\alpha(H)}{dH} &= R^{-1}\sum_{\beta = 1}^{N_s}\left( r_{\alpha\beta}(H) p_\alpha(H) - r_{\beta\alpha}(H)p_\beta(H)\right). 
\label{MHeq}
\end{align}
Integrating these equations with the initial condition $p_1(H=H_i) = 1$ yields the probabilities $p_\alpha(H)$ from which we can then compute the magnetization as a function of $H$,
\begin{align}
m(H) = \sum_\alpha m_\alpha p_\alpha (H),
\end{align}
where $m_\alpha$ denotes magnetic moment of a grain in state $\sigma_\alpha$.

Calculating the rate constant $r_{\alpha\beta}$ as a function of $H$ requires that we determine the location of the minimum energy states and the saddle point located on the boundaries separating their basins of attraction for each field value. For this simple example, locating the energy minima is straightforward. How best to determine the location of the saddle points is less obvious. One technique is to compute the MEP that connects the two minima $\sigma_\alpha \leftrightarrow \sigma_\beta$. This may be done numerically by discretizing an initial guess of the MEP and allowing the points to relax until the derivatives of the energy perpendicular to the tangent line at each point of the path are zero. Two methods that successfully implement this scheme are the Nudged Elastic Band (NEB) method and the string method.\cite{henkelmana,weinan1,weinan2} In the present work, we have used a variant of the string method to calculate the MEPs. MEPs for both the strong coupling ($I=2.0\times10^{-3}\:\mathrm{J/m^2}$) and the weak coupling ($I=0.5\times10^{-3}\:\mathrm{J/m^2}$) regimes are shown in Figs.~\ref{Almudallal_fig01} and \ref{Almudallal_fig02} for both $H$=0 and $H$=4\:kOe. It should be noted that not every pair of energy minima are directly connected by an MEP. For such cases $r_{\alpha\beta} = 0$. 

Parametric plots of the energy $E(\theta_a,\theta_b)$ along the MEP are shown in Figs.~\ref{Almudallal_fig03_a} and \ref{Almudallal_fig03_b}  for $I=2.0\times 10^{-3}\:\mathrm{J/m^2}$ and in Figs.~\ref{Almudallal_fig04_a} - \ref{Almudallal_fig04_d}  for $I=0.5\times 10^{-3}\:\mathrm{J/m^2}$. The figures also show parametric plots for the energy along the initial path. For the strong coupling regime ($I=2.0\times 10^{-3}\:\mathrm{J/m^2}$), the initial path used is given by $(\theta,\theta)$, while in the weak coupling regime ($I=0.5\times 10^{-3}\:\mathrm{J/m^2}$), where there are up to four MEPs,  the initial paths used were $(\theta,-\pi/2)$, $(\pi/2,\theta)$, $(-\pi/2,\theta)$, and $(\theta,\pi/2)$ where $-\pi/2 \le \theta \le \pi/2$. The saddle point is located at the point of the peak energy on the MEP. 

\begin{figure}[ht]
\subfigure{
\centering\includegraphics[clip=true, trim=0 0 0 0, height=5.0cm]{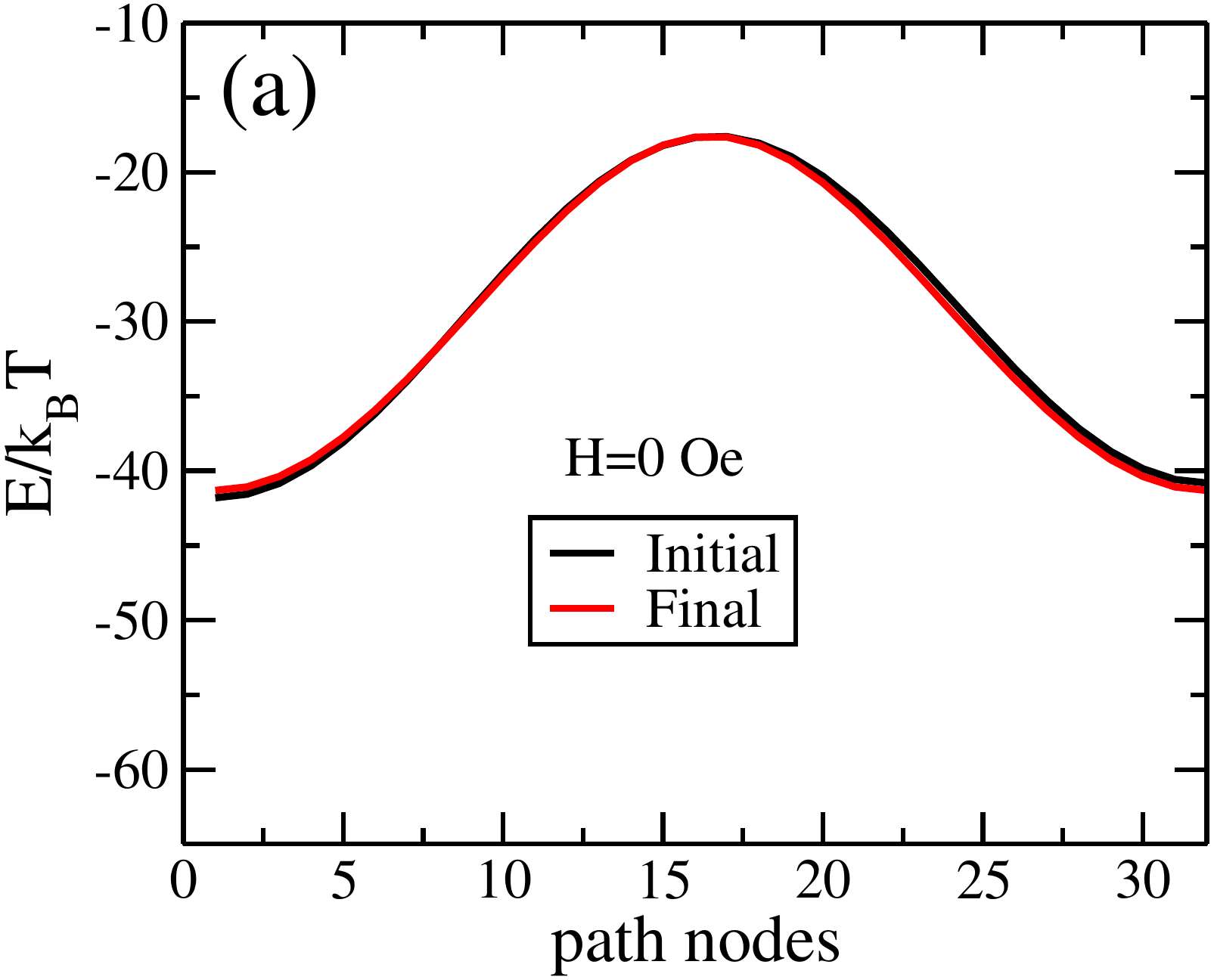}
\label{Almudallal_fig03_a}
}
\subfigure{
\centering\includegraphics[clip=true, trim=0 0 0 0, height=5.0cm]{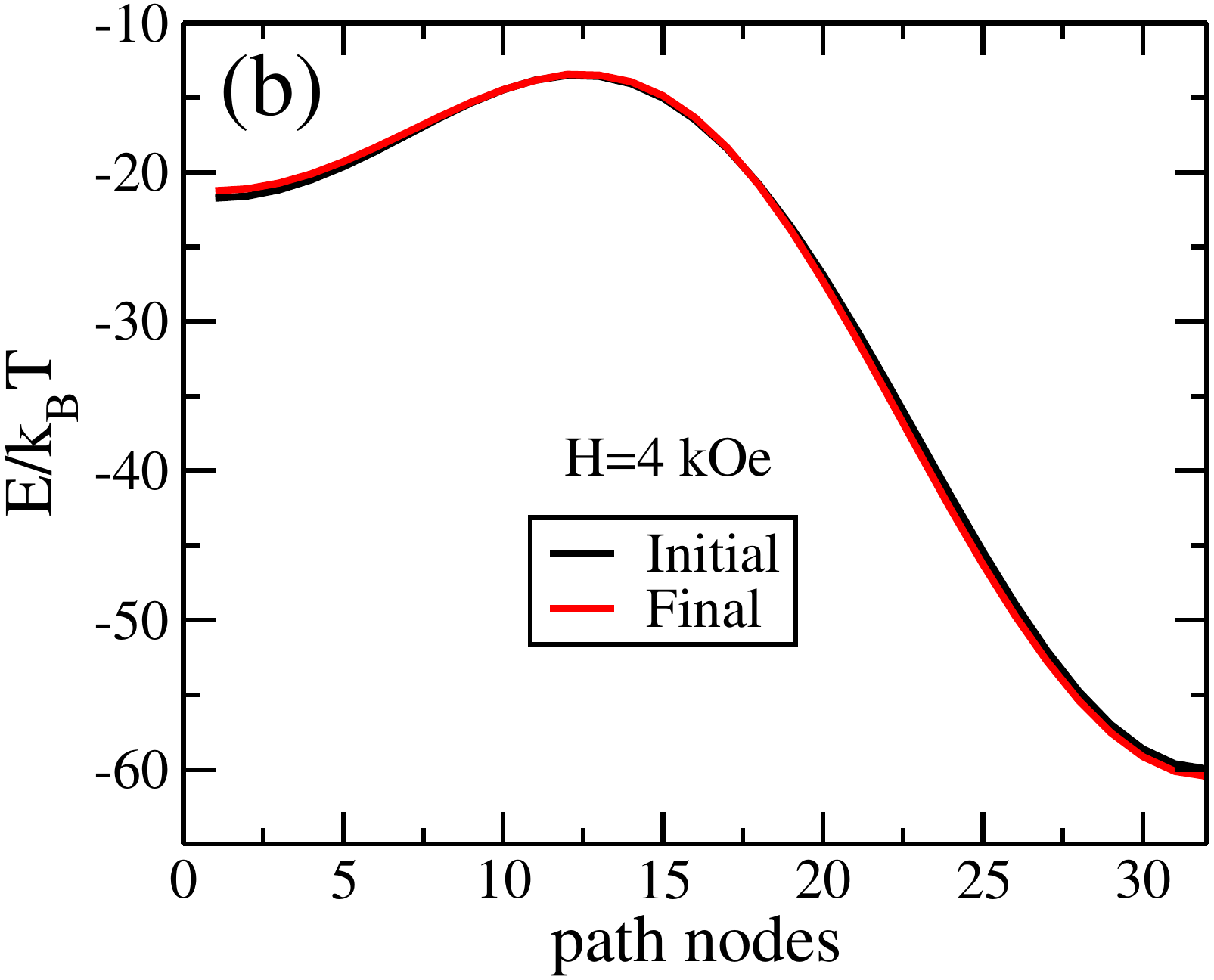}
\label{Almudallal_fig03_b}
}
\caption{A plot of the energy along the length of the parametric MEP (red line) and the initial path used  to generate it (black line) for $I$=2.0$\times 10^{-3}$\:J/m$^2$ (a) $H$=0 and  (b) $H$=4\:kOe.}
\label{Almudallal_fig03}
\end{figure}

\begin{figure}[ht]
\subfigure{
\centering\includegraphics[clip=true, trim=0 0 0 0, height=5.0cm]{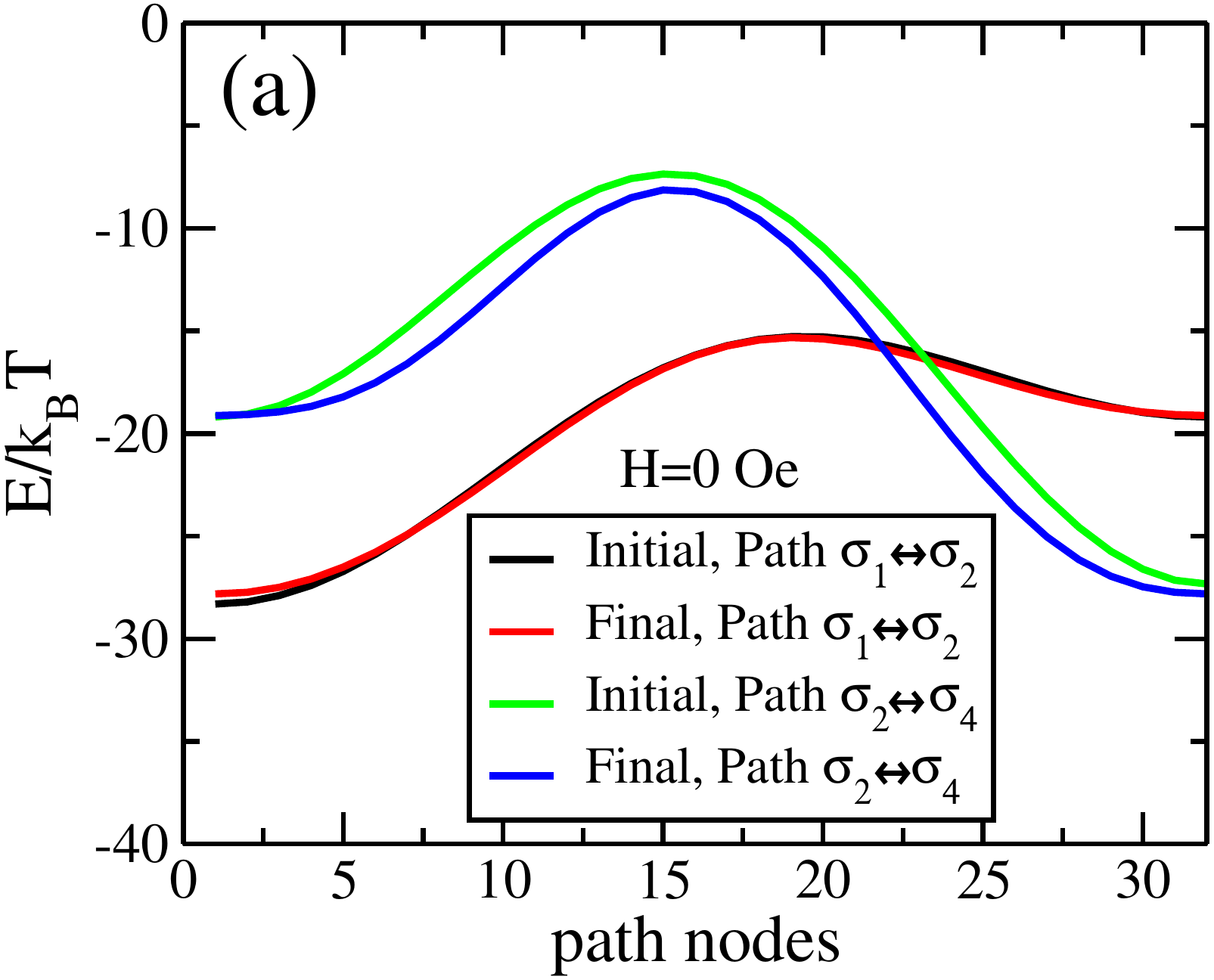}
\label{Almudallal_fig04_a}
}
\subfigure{
\centering\includegraphics[clip=true, trim=0 0 0 0, height=5.0cm]{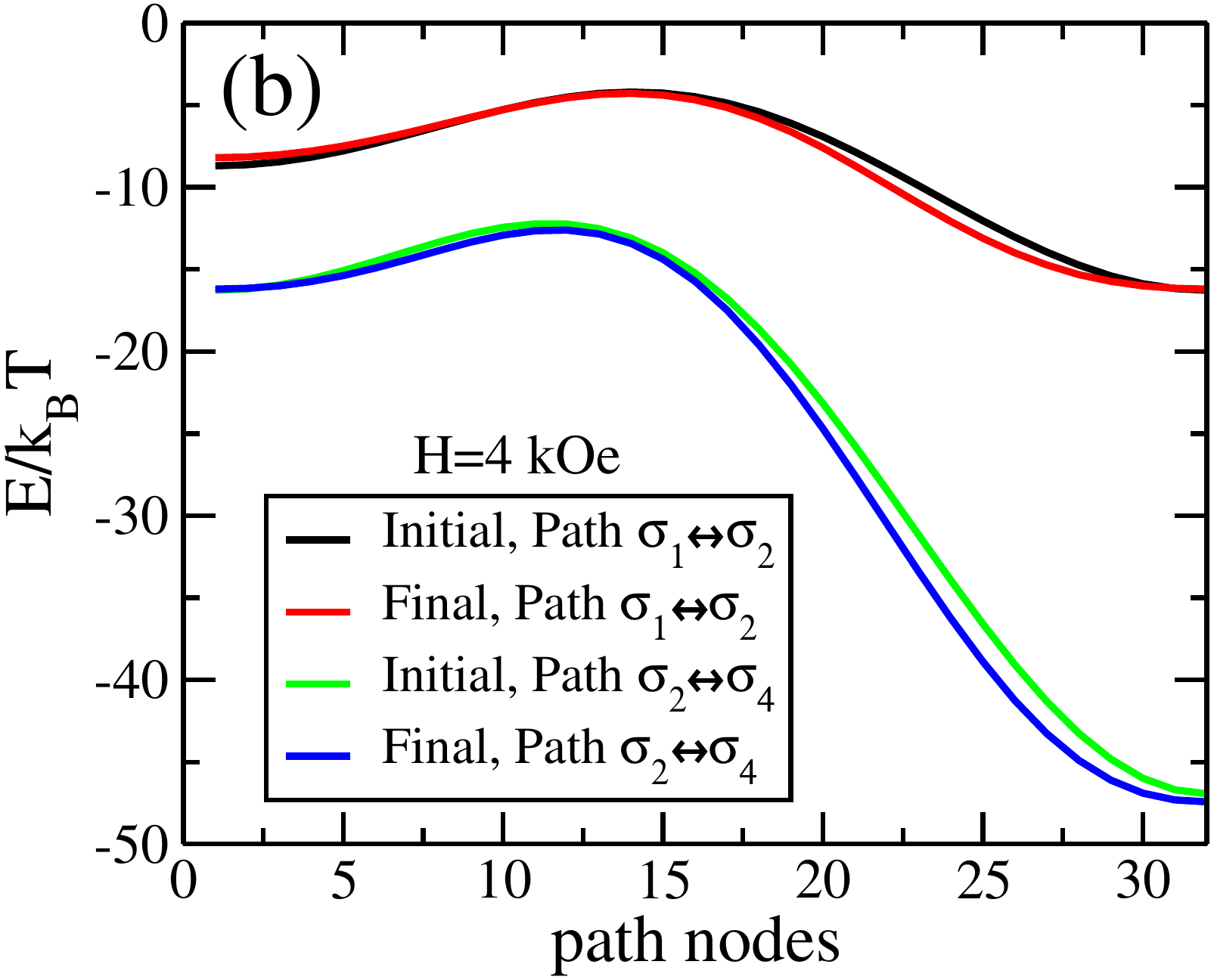}
\label{Almudallal_fig04_b}
}
\subfigure{
\centering\includegraphics[clip=true, trim=0 0 0 0, height=5.0cm]{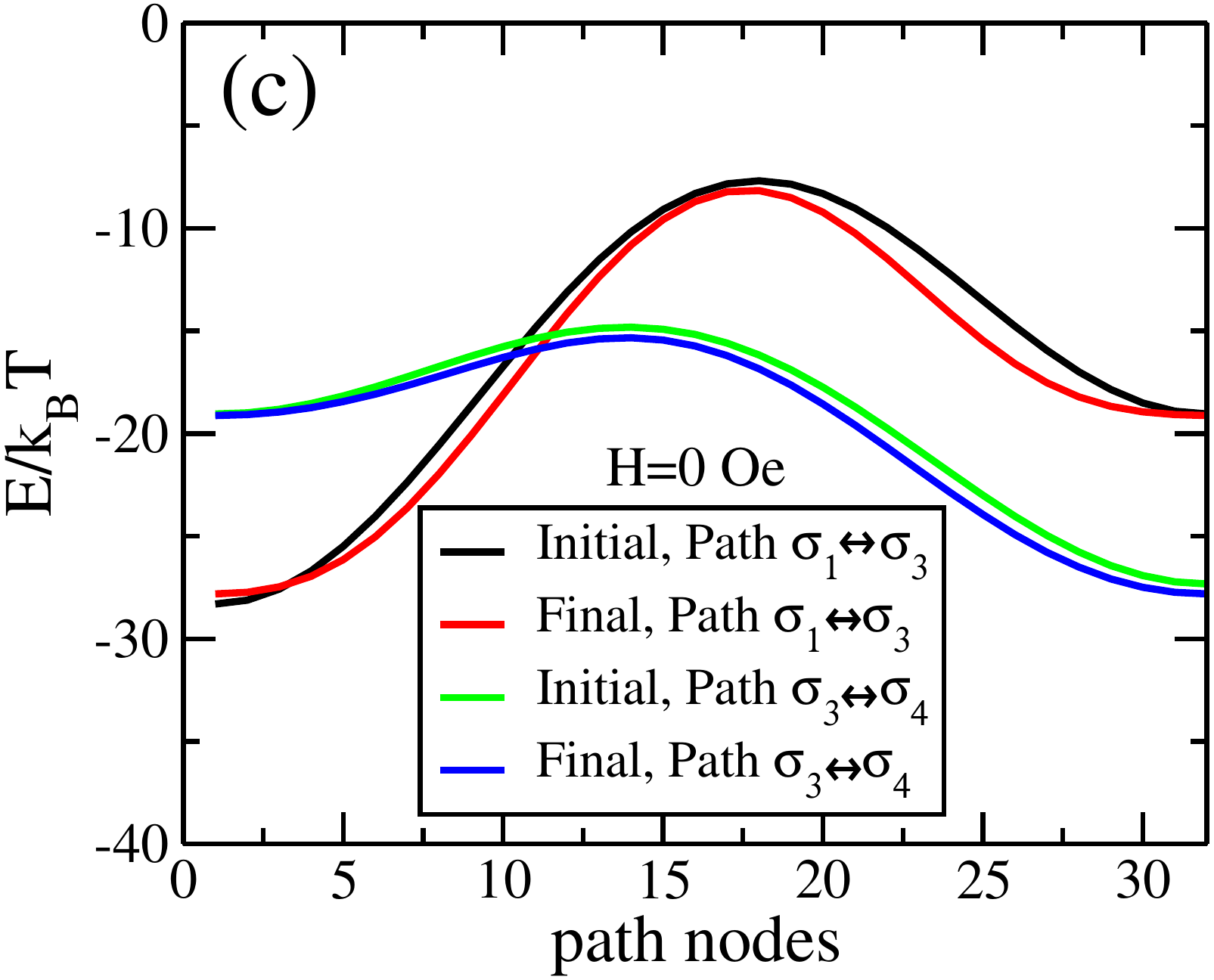}
\label{Almudallal_fig04_c}
}
\subfigure{
\centering\includegraphics[clip=true, trim=0 0 0 0, height=5.0cm]{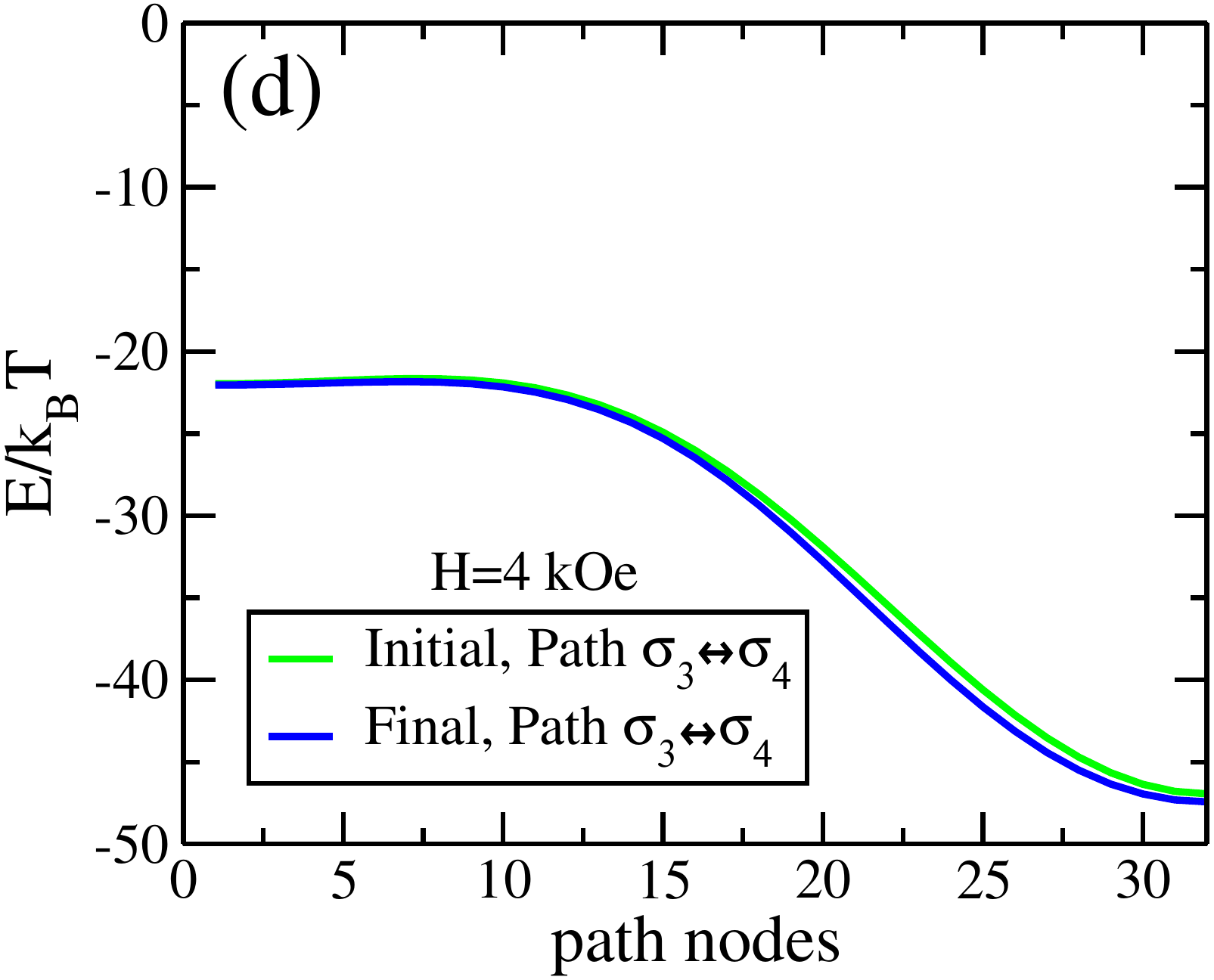}
\label{Almudallal_fig04_d}
}
\caption{A plot of the energy along the length of the four parametric MEPs (red and blue lines) and the initial paths used  to generate them (black  and green lines) for $I$=0.5$\times 10^{-3}$\:J/m$^2$ (a) $H$=0 and (b) $H$=4\:kOe.}
\label{Almudallal_fig04}
\end{figure}

The integration of Eq.~\eqref{MHeq} proceeds as follows, the minimum energy states for several values of $H$ over the range $-5\:\mathrm{kOe} \le \mu_0 H \le 2\:\mathrm{kOe}$ and the MEPs joining them are determined. The saddle points are located at the point of maximum energy on the MEP. Once the minimum energy states and the saddle points have been determined, the non-zero rate constants $r_{\alpha\beta} (H)$ are calculated using the expression given by Eqs.~\eqref{ArrheniusNeel} and \eqref{attemptFrequency} at these selected values of $H$. The rate coefficients for values of $H$ intermediate between these discrete values are then determined by interpolation. The rate equation~\eqref{MHeq} is then be solved numerically using Mathematica. 

The range of sweep rates chosen corresponds approximately to time scales involved in experimental $MH$ hysteresis loop measurements, $R \sim$ $10^3$\:Oe/s, to magnetic recording rates, $R \sim$  $10^{10}$\:Oe/s.\cite{plumer4} Figure~\ref{Almudallal_fig05} shows the calculated $MH$ hysteresis loops at $T$=300\:K with $\alpha_0 = 0.1$ for different exchange coupling values $I=2.0,\:1.5,\:1.0,\:0.5,\:0.25 \; \mathrm{and}\;  0.1\times 10^{-3}\:\mathrm{J/m}^2$, respectively. $MH$ hysteresis loops are calculated at all the different sweep rates, but only the range  $10^{5}\:\mathrm{Oe/s} \le R \le10^{10}\:\mathrm{Oe/s}$ are shown in these figures.  
 
\begin{figure}[ht]
\subfigure{
\centering\includegraphics[clip=true, trim=0 0 0 0, height=5.0cm]{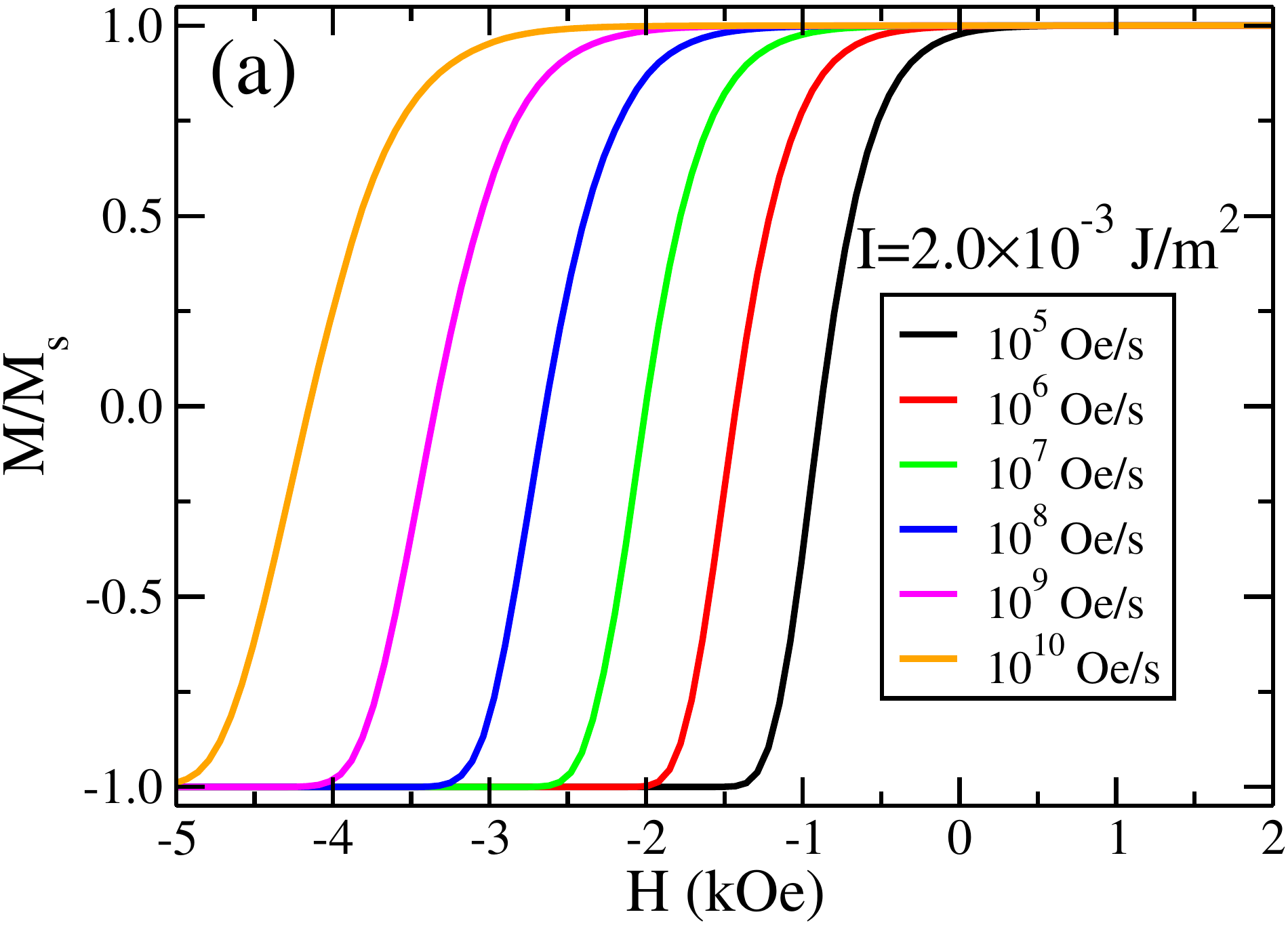}
\label{Almudallal_fig05_a}
}
\subfigure{
\centering\includegraphics[clip=true, trim=0 0 0 0, height=5.0cm]{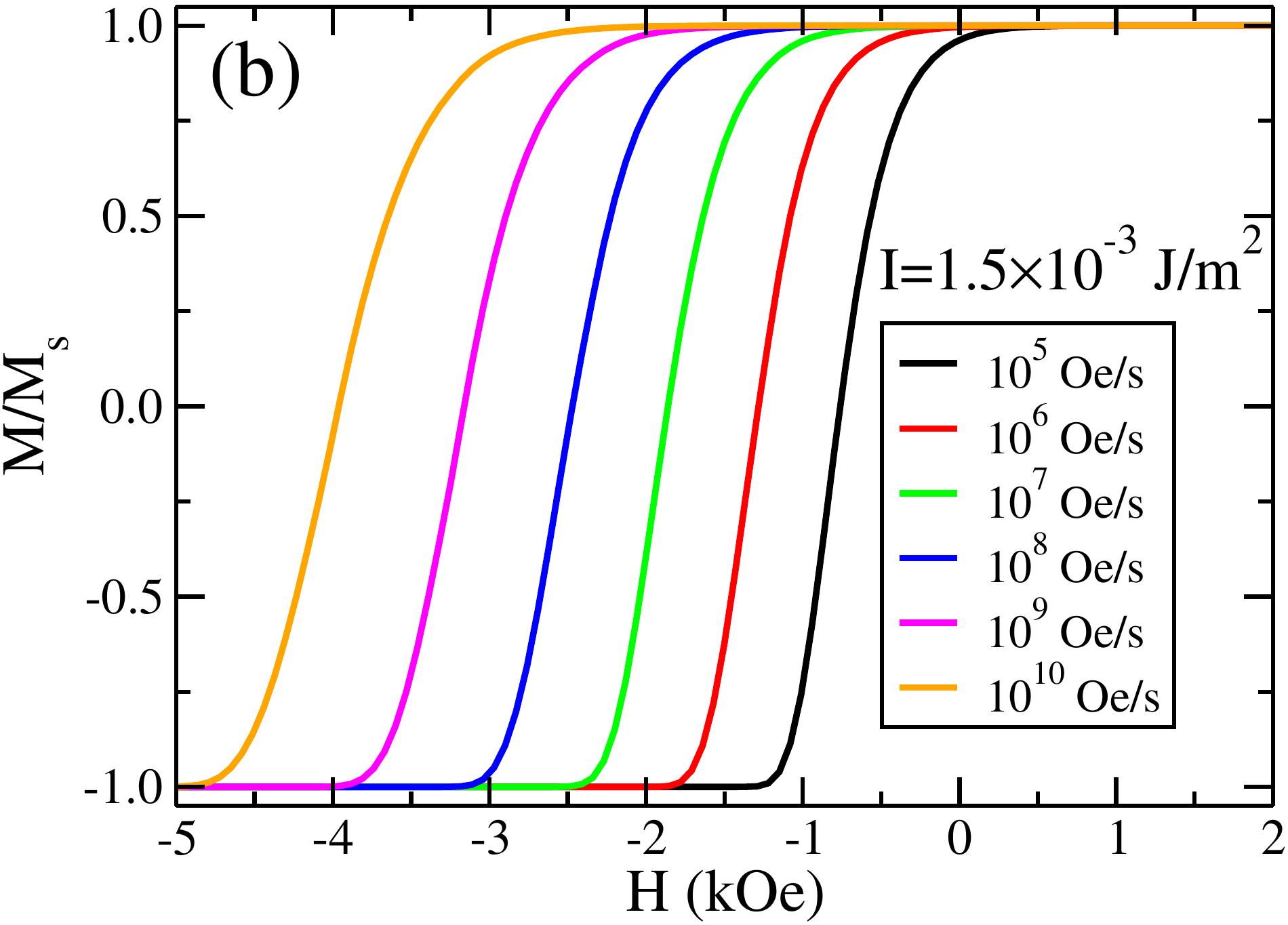}
\label{Almudallal_fig05_b}
}
\subfigure{
\centering\includegraphics[clip=true, trim=0 0 0 0, height=5.0cm]{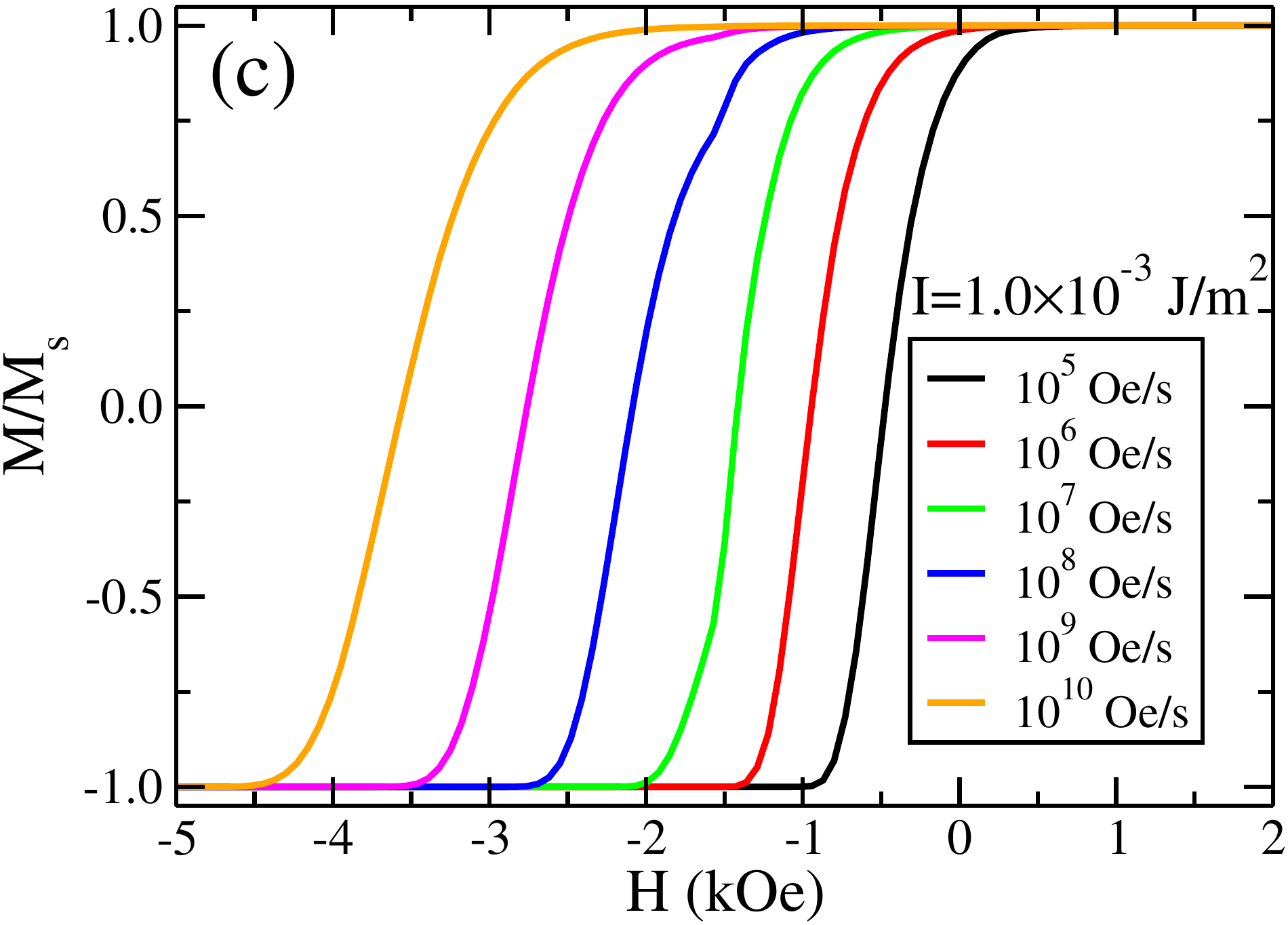}
\label{Almudallal_fig05_c}
}
\subfigure{
\centering\includegraphics[clip=true, trim=0 0 0 0, height=5.0cm]{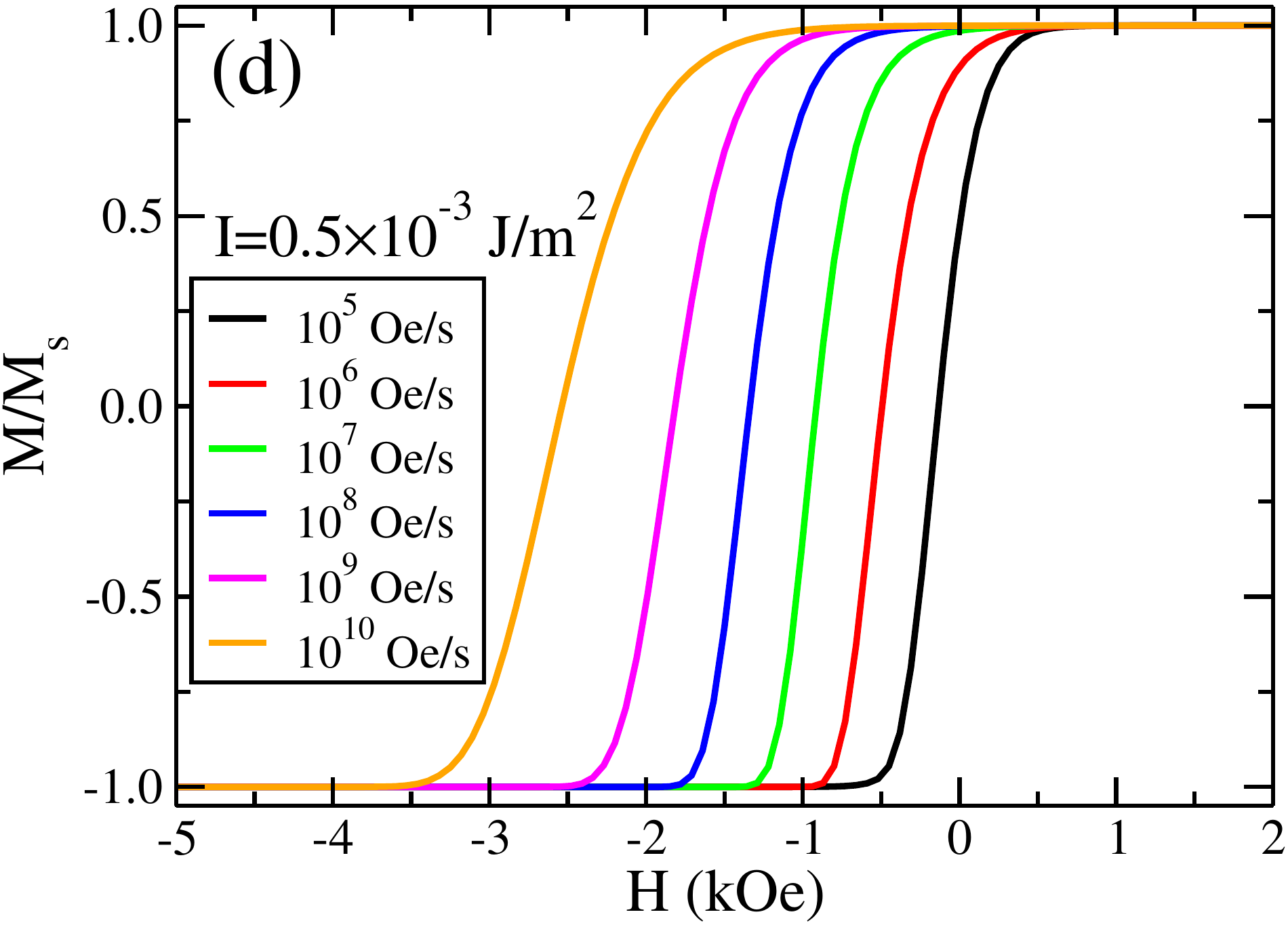}
\label{Almudallal_fig05_d}
}
\subfigure{
\centering\includegraphics[clip=true, trim=0 0 0 0, height=5.0cm]{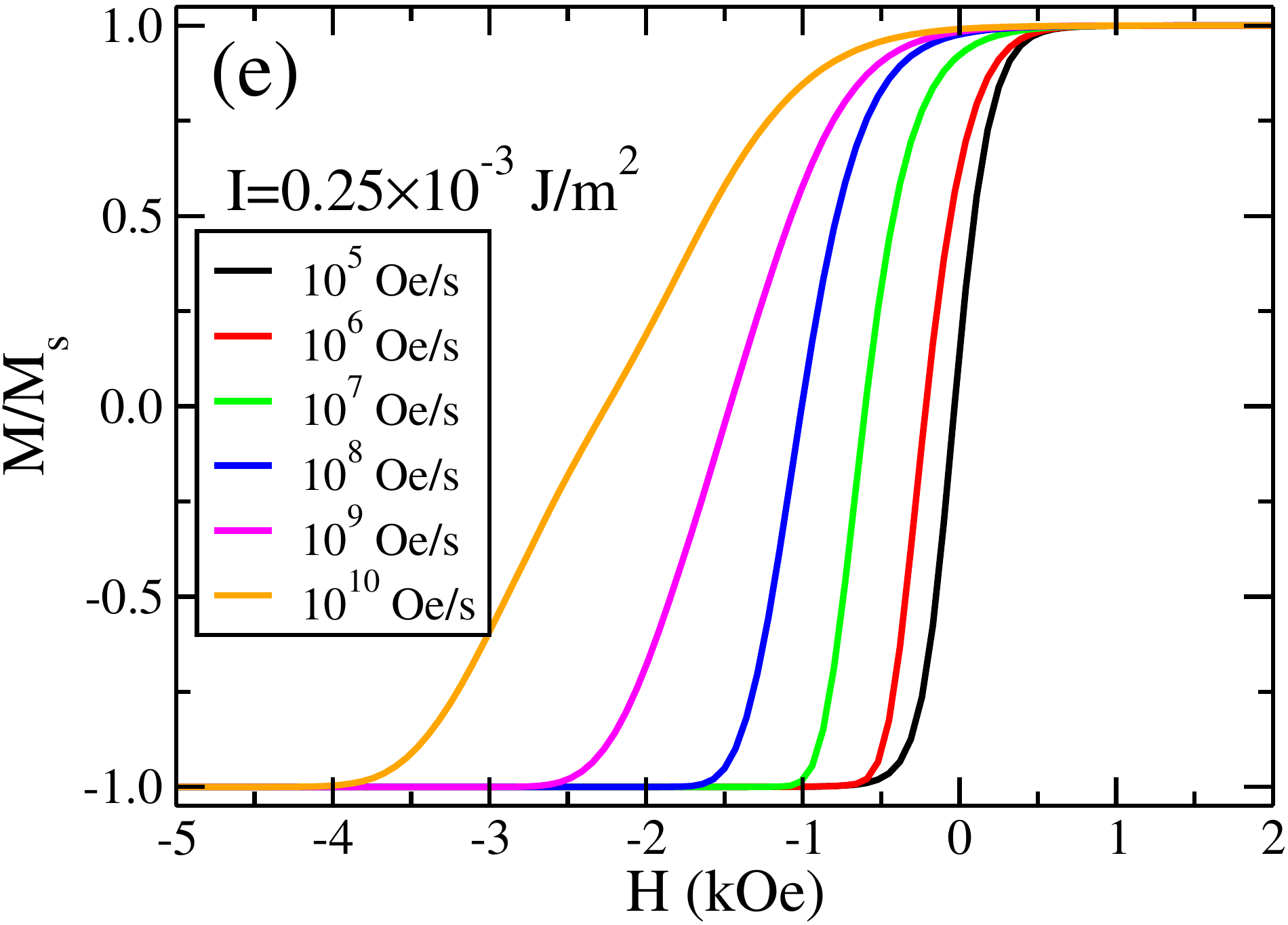}
\label{Almudallal_fig05_e}
}
\subfigure{
\centering\includegraphics[clip=true, trim=0 0 0 0, height=5.0cm]{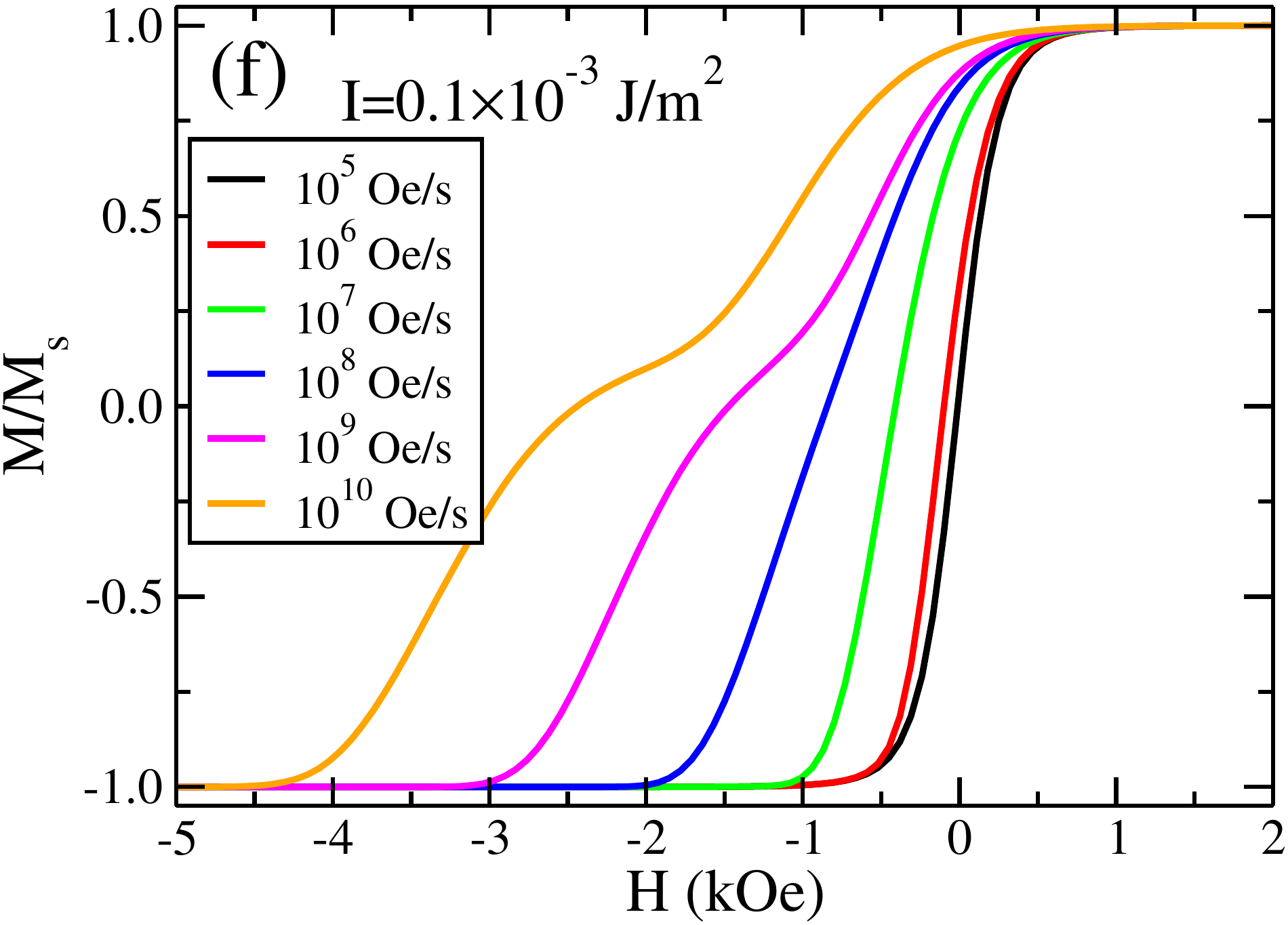}
\label{Almudallal_fig05_f}
}
\caption{$MH$ hysteresis loops calculated at $T$=300\:K by direct integration of the rate equations for different sweep rates $R$: (a) $I$=2.0$\times 10^{-3}$\:J/m$^2$, (b) $I$=1.5$\times 10^{-3}$\:J/m$^2$, (c) $I$=1.0$\times 10^{-3}$\:J/m$^2$, (d) $I$=0.5$\times 10^{-3}$\:J/m$^2$, (e) $I$=0.25$\times 10^{-3}$\:J/m$^2$, and (f) $I$=0.1$\times 10^{-3}$\:J/m$^2$. }
\label{Almudallal_fig05}
\end{figure}

From these results, the expected trend of the coercivity $H_c$ decreasing at slower sweep rates can be observed. In addition, there is little difference between the strong coupling cases of $I$=2.0$\times 10^{-3}$\:J/m$^2$ and $I$=1.5$\times 10^{-3}$\:J/m$^2$. Moderate coupling $I$=1.0$\times 10^{-3}$\:J/m$^2$ and $I$=0.5$\times 10^{-3}$\:J/m$^2$ represents a crossover regime between the two grains acting as a single grain, and the two grains responding quasi-independently.  Here the hysteresis loops are quite sensitive to the coupling $I$.  Weak coupling is clear in the case of $I$ = 0.1$\times 10^{-3}$\:J/m$^2$ at the fast sweep rate, where the plateau indicates that the soft grain switches first.  

From the hysteresis loops, we can extract the nucleation field $H_n=H(M/M_s=0.95)$, the coercive field $H_c=H(M/M_s=0.0)$, and the saturation field $H_s=H(M/M_s=-0.95)$.\cite{plumer4} Figure~\ref{Almudallal_fig06} shows these extracted values of $H_n$, $H_c$, and $H_s$ as a function of $I$ for different sweep rates. Although the nucleation field exhibits monotonic decrease with increasing $I$ and $R$, both $H_c$ and $H_s$ show clear minima at weak to moderate coupling values in the cases of the faster sweep rates.  

\begin{figure}[ht]
\subfigure{
\centering\includegraphics[clip=true, trim=0 0 0 0, height=5.0cm]{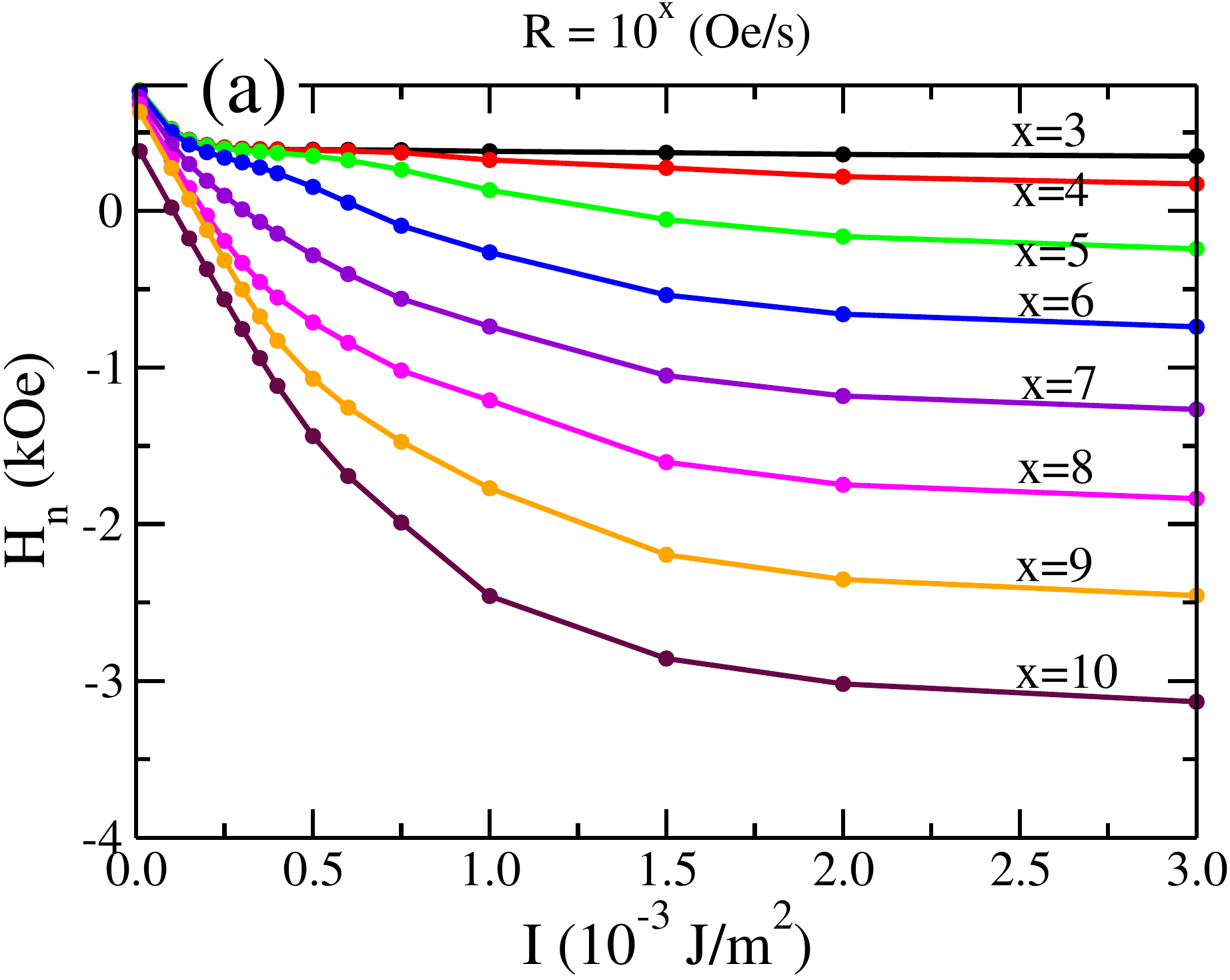}
\label{Almudallal_fig06_a}
}
\subfigure{
\centering\includegraphics[clip=true, trim=0 0 0 0, height=5.0cm]{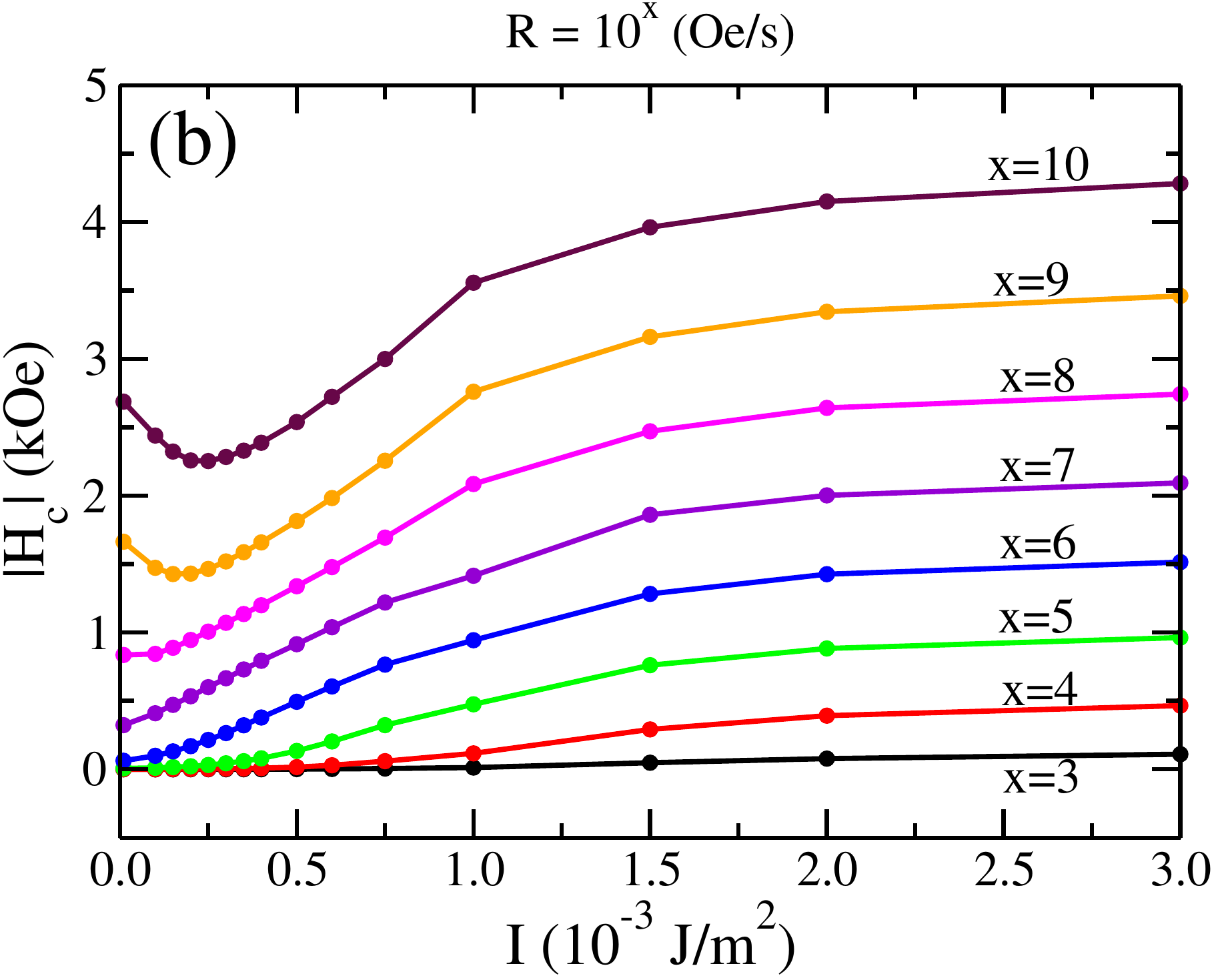}
\label{Almudallal_fig06_b}
}
\subfigure{
\centering\includegraphics[clip=true, trim=0 0 0 0, height=5.0cm]{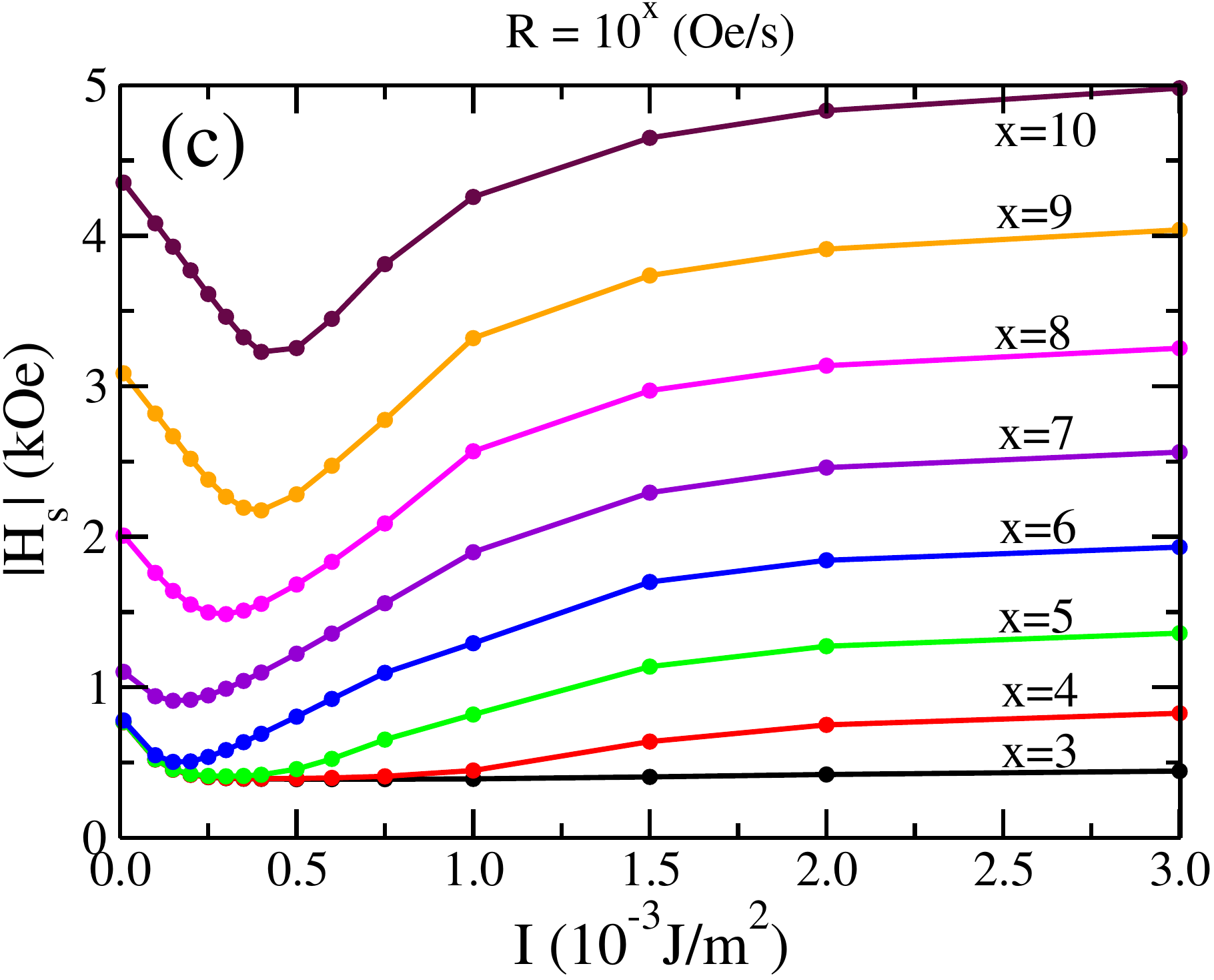}
\label{Almudallal_fig06_c}
}
\caption{(a) The nucleation field $H_n$, (b) coercivity $H_c$, and (c) saturation field $H_s$ extracted from Fig.~\ref{Almudallal_fig05} as a function of $I$ at different sweep rates.}
\label{Almudallal_fig06}
\end{figure}

\section{Comparison with LLG}
\label{llg}

As mentioned above and discussed in more detail in the Appendix, the calculation of the rate coefficients follows from the FPE, which can be derived from the stochastic LLG equation.\cite{brown79,gardiner,langer} In fact the derivation of the FPE from the stochastic LLG equation imposes non-trivial requirements on the integration schemes that can be used to solve the stochastic LLG equation.  It is therefore interesting to compare the $MH$ hysteresis loops obtained from stochastic LLG and those obtained in Sec.~\ref{MHloops}. Previous comparisons for interacting grains where the rate equations have been solved using both Kinetic Monte Carlo (KMC)\cite{fal1,plumer4} and stochastic LLG have shown good agreement between the two approaches over a limited range of sweep rates ($10^8\:\mathrm{Oe/s} \le R \le 10^{10}\:\mathrm{Oe/s}$). The range of sweep rates over which we might expect good agreement between the two approaches is limited by the fact that LLG results are only accessible within a reasonable amount of simulation time for $R \ge 10^8\:\mathrm{Oe/s}$ while for $R > 10^{10}\: \mathrm{Oe/s}$, the Arrhenius-N\'{e}el expression for the rate coefficient, that serves as a basis for the KMC algorithm, breaks down, as it does not fully capture the spin dynamics of the reversal process. 

$MH$ hysteresis loops obtained from LLG simulations for a system of $16\times16$ non-interacting, exchange coupled dual layer grains using the same parameters detailed in Sec.~\ref{energyLandscapes} are presented in Figs.~\ref{Almudallal_fig07_a} -  \ref{Almudallal_fig07_c} together with loops obtained by the MEP method. The time step used was $2$\:ps and the integration was performed using the Runge-Kutta fourth order method based on a quaternion representation of the rotations with the damping parameter set at $\alpha_0=0.1$. The simulations performed at $T=300\:\mathrm{K}$.
The comparison shown in Fig.~\ref{Almudallal_fig07_a} for $I$=2.0$\times 10^{-3}$\:J/m$^2$,  Fig.~\ref{Almudallal_fig07_b} for $I$=0.5$\times 10^{-3}$\:J/m$^2$, and Fig.~\ref{Almudallal_fig07_c} for $I$=0.1$\times 10^{-3}$\:J/m$^2$ indicates a very good agreement between the two methods at all sweep rates.    

\begin{figure}[ht]
\subfigure{
\centering\includegraphics[clip=true, trim=0 0 0 0, height=5.0cm]{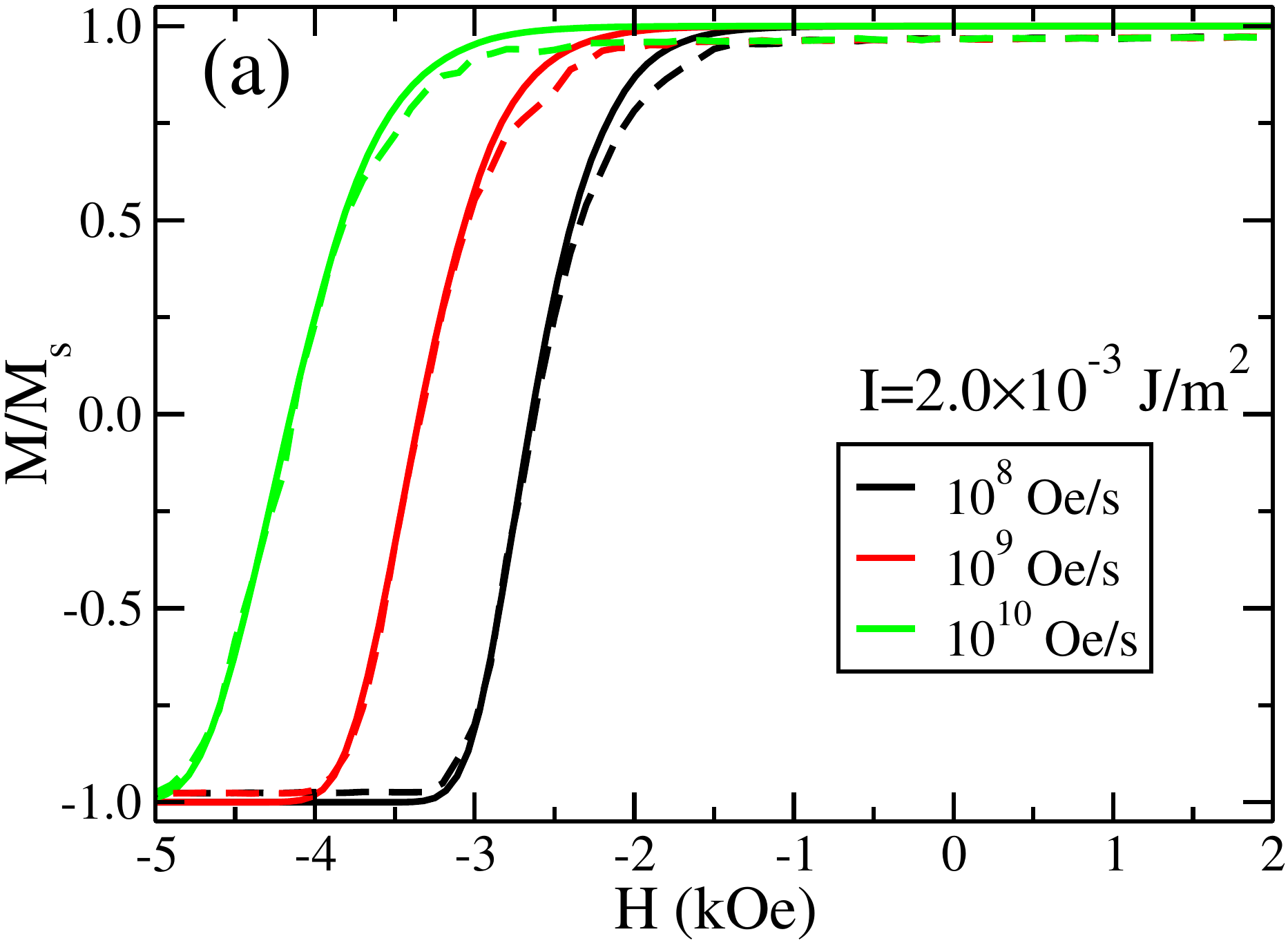}
\label{Almudallal_fig07_a}
}
\subfigure{
\centering\includegraphics[clip=true, trim=0 0 0 0, height=5.0cm]{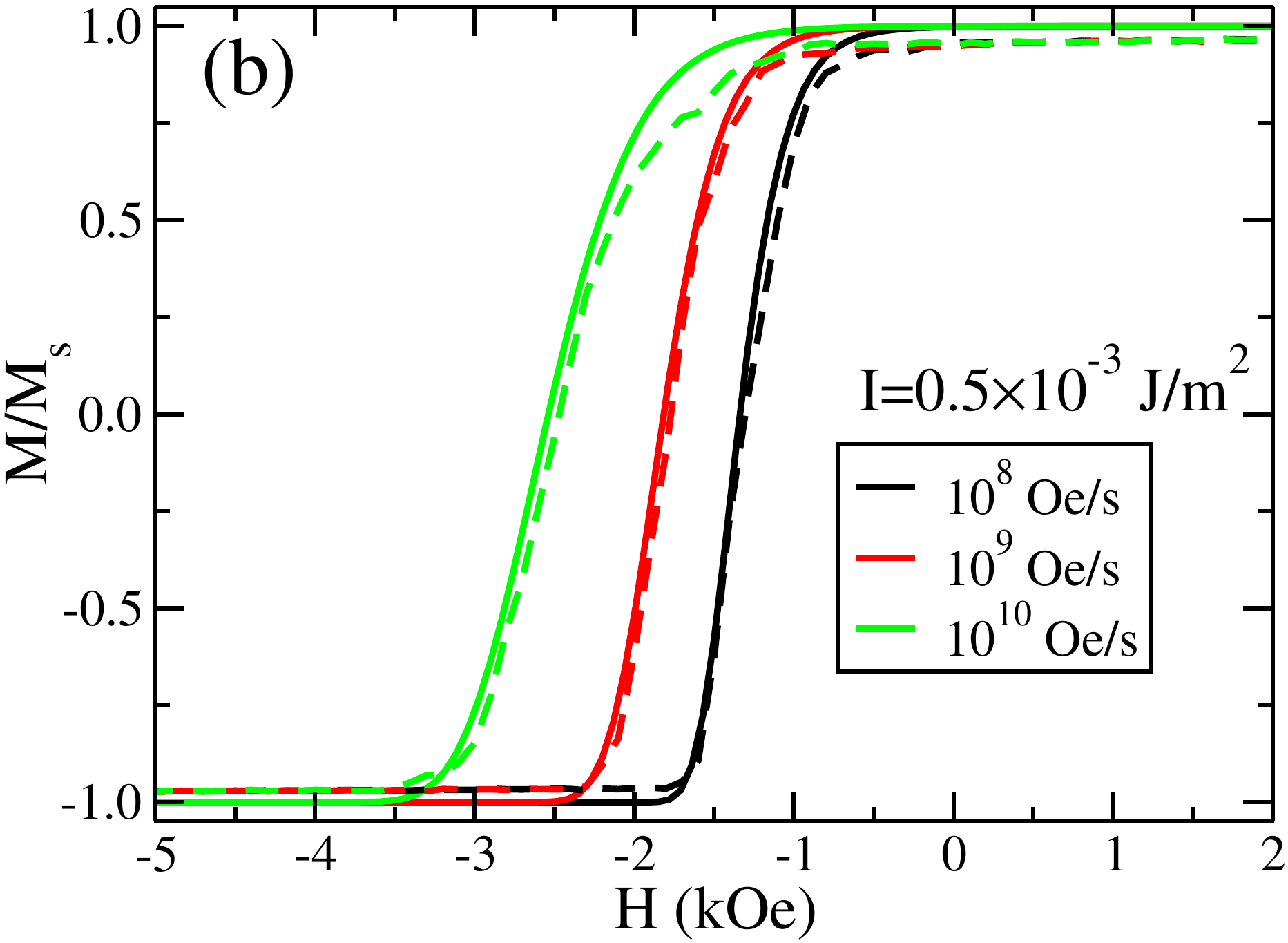}
\label{Almudallal_fig07_b}
}
\subfigure{
\centering\includegraphics[clip=true, trim=0 0 0 0, height=5.0cm]{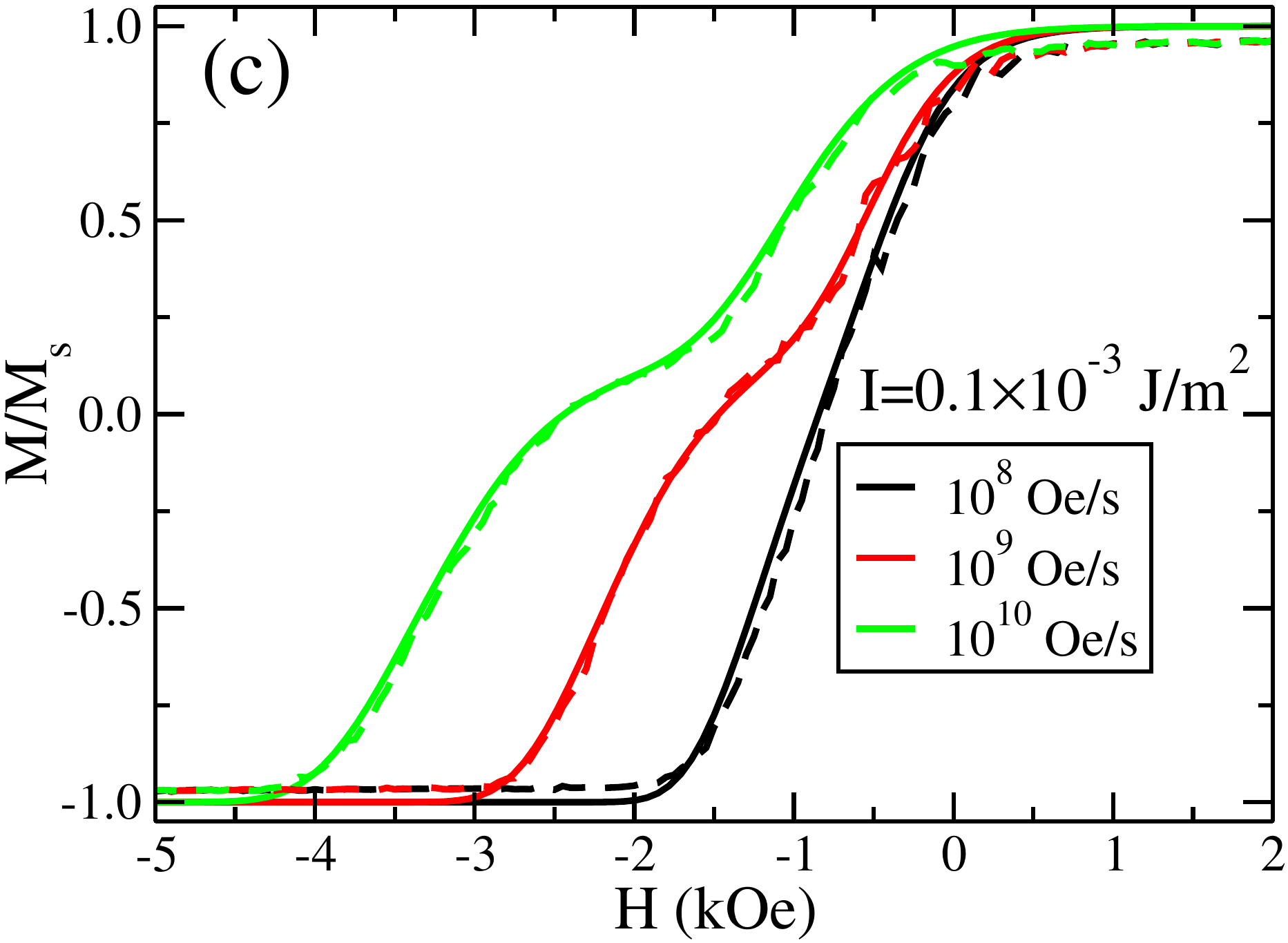}
\label{Almudallal_fig07_c}
}
\caption{Comparison of $MH$ hysteresis loops from the MEP method (solid curves) and stochastic LLG (dashed curves) at different sweep rates for (a) $I$=2.0$\times 10^{-3}$\:J/m$^2$, (b) $I$=0.5$\times 10^{-3}$\:J/m$^2$, and (c) $I$=0.1$\times 10^{-3}$\:J/m$^2$.}
\end{figure}

\section{Figure of Merit for ECC media}
\label{results}

The benefit of coupling hard and soft layers can be quantified in a figure of merit (FOM), which is the ratio of a measure of the thermal stability and the field required to switch the grain magnetization.\cite{victora,kapoor,wang,suess,richter,choe} This can be defined as the ratio between the energy barrier $E_B$ (thermal stability) and saturation field (switching energy) at a particular sweep rate as, 

\begin{equation}
\text{FOM} = \frac{E_B}{\mu_0 H_s (M_a v_a + M_b v_b)}.
\label{fom_eq}
\end{equation}
For strong coupling, $E_B$ is given by the zero field energy barrier between the minimum energy of state $\sigma_1$ and the saddle point along the path to the minimum energy of state $\sigma_4$, while for weak coupling, $E_B$ is the zero field energy barrier between the minimum energy of state $\sigma_1$ and the saddle point along the path to the minimum energy of state $\sigma_2$. A larger FOM is the goal for ECC-type media. The results shown in Fig.~\ref{Almudallal_fig06_c} indicate that increasing $I$, for small $I$, will decrease the saturation field which makes switching the magnetic moment easier. On the other hand, increasing $I$ will increase the energy barrier which enhances the thermal stability (not shown).  Fig.~\ref{Almudallal_fig08} shows the FOM at three sweep rates ($R=10^6$, $10^8$, and $10^{10}$\:Oe/s), and the optimal value of $I$ can be easily obtained from the graph: $I_\mathrm{op}$($10^6$\:Oe/s)$\sim$0.2$\times 10^{-3}$\:J/m$^2$,\: $I_\mathrm{op}$($10^8$\:Oe/s)$\sim$0.35$\times 10^{-3}$\:J/m$^2$, and  $I_\mathrm{op}$($10^{10}$\:Oe/s)$\sim$0.50$\times 10^{-3}$\:J/m$^2$.  These results suggest that weak to moderate coupling is preferred and that there is a strong dependence on sweep rate.  Large FOM values at smaller sweep rates may not be realized at larger sweep rates, and optimal coupling strengths estimated on the basis of experimental $MH$ hysteresis loops obtained at slow sweep rates may thus not be the optimal value at recording time scales.

\begin{figure}[ht]
\centering\includegraphics[clip=true, trim=0 0 0 0, height=5.0cm]{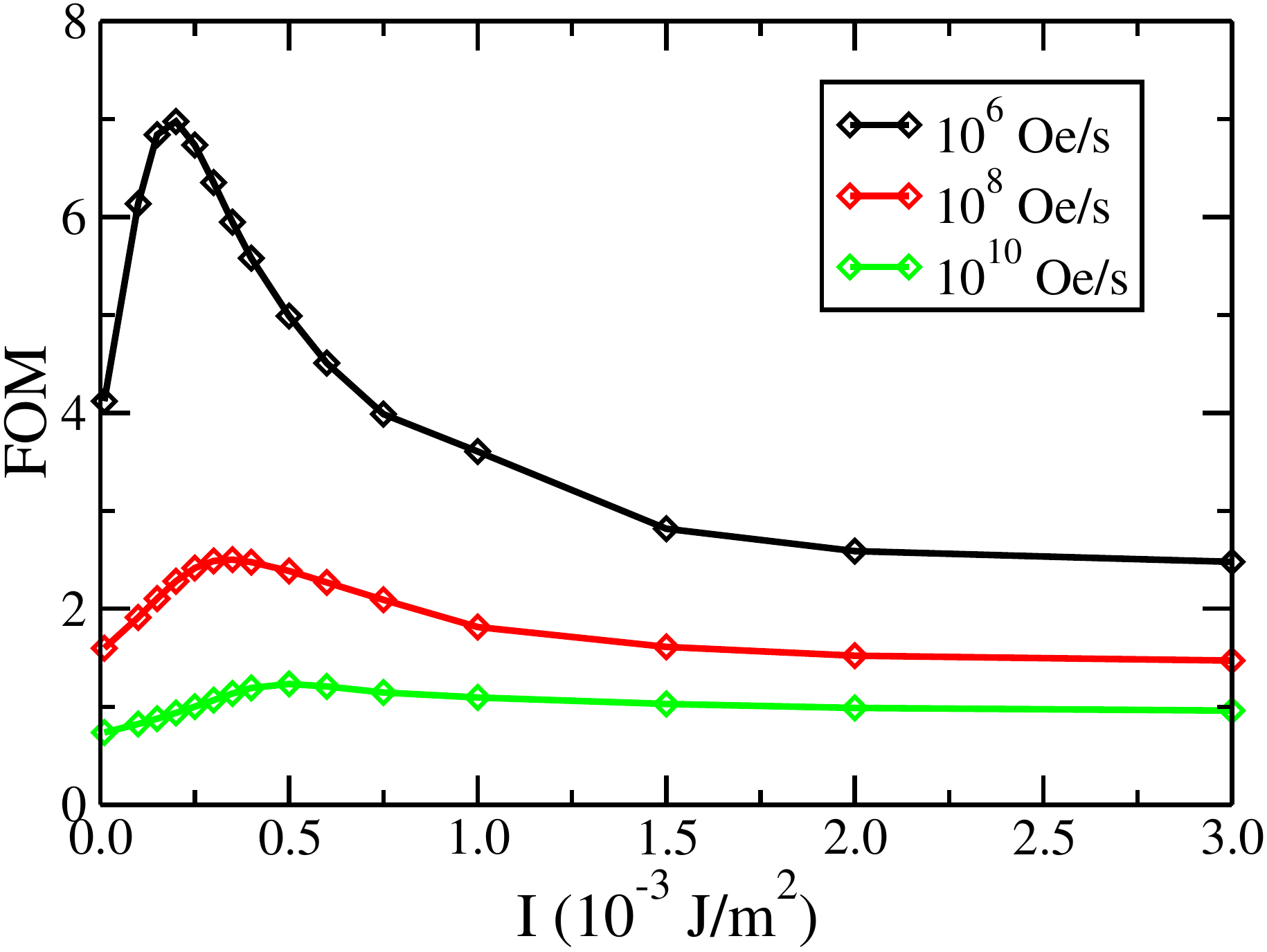}
\caption{Figure of merit calculated by Eq.~\eqref{fom_eq} for three sweep rates ($10^6$, $10^8$, and $10^{10}$\:Oe/s).}
\label{Almudallal_fig08}
\end{figure}

\section{Approximation Schemes}
\label{am-sec}

In this section we describe some approximation schemes which allow simplification of the rate equations used in Sec.~\ref{MHloops}, not only making calculations less onerous but also, in certain cases, allowing for analytical solutions. Comparisons with the exact rate equations show that for certain regions of parameter space, these approximation methods are surprisingly accurate and can provide insight into the complex nature of the reversal process in ECC media. 

\subsection{Direct Path Approximation} 

Figures \ref{Almudallal_fig03} and \ref{Almudallal_fig04} show that the energy calculated along the paths used as an initial guess in the determination of  the MEP are in fact very close to those given by the MEP for both the strong coupling case ($I=2.0\times 10^{-3}$\:J/m$^2$) and the weak coupling case ($I=0.5\times 10^{-3}$\:J/m$^2$). This suggests that, in the strong coupling case, a reasonably good approximation to the rate coefficients can be found by replacing the MEP with the direct path $\theta_1 = \theta_2 = \theta$. For this path  the energy can be written as, 
\begin{align}
E &= -\left(K_a v_a + K_b v_b \right)\sin^2\theta - \mu_0 H \left(M_a v_a + M_b v_b\right) \sin\theta -IA.
\end{align}
This expression for the energy is of the SW form and hence the expressions for the attempt frequency and energy barrier can be found analytically using the expressions  in Brown's classic paper,\cite{brown}
\begin{align}
f_{\alpha\beta} &= \sqrt{\dfrac{K_T v}{\pi k_B T}} \left(\dfrac{\alpha_0\:\gamma}{1+\alpha_0^2} \right) \left(1-\dfrac{H^2}{{H_K}^2}\right) \left(H_K\pm H\right),\label{brownEqs1}\\
\Delta E_{\alpha\beta} &=-K_Tv\left(1\mp \dfrac{H}{H_K}\right)^2,\label{brownEqs2}
\end{align}
where $H_K = 2K_T/M_T$, $K_T = K_a + K_b$, and $M_T = M_a + M_b$. In Fig.~\ref{Almudallal_fig09}(a),  we show a comparison of the $MH$ hysteresis loops calculated using the rate coefficients calculated using the MEP method and the direct path approximation, with $\alpha_0$ = 0.1 and $I$=2.0$\times 10^{-3}$\:J/m$^2$ for several sweep rates. 

Similarly in the weak coupling case, we can replace the four MEPs that link the minima $\{\sigma_1 \leftrightarrow \sigma_2,\sigma_2 \leftrightarrow \sigma_4, \sigma_1 \leftrightarrow \sigma_3,  \sigma_3 \leftrightarrow \sigma_4, \}$ by the paths $\{\theta_a ,\theta_b\} \in\left\{ \{\theta,-\pi/2\},\{\pi/2,\theta\}, \{-\pi/2,\theta\},\{\theta, \pi/2\}\right\}$. It is straightforward to show that along each of the paths, the energy will be of the SW form and the rates may be calculated using Eqs.~\eqref{brownEqs1} and~\eqref{brownEqs2}. In Fig.~\ref{Almudallal_fig09}(b), we show a comparison of the $MH$ hysteresis loops calculated by the MEP method and the direct path approximation with $I$=0.5$\times 10^{-3}$\:J/m$^2$, for several sweep rates. The coercive field is shown as a function of sweep rate for $I$=2.0,\:0.5, and 0.1$\times 10^{-3}$\:J/m$^2$ in Fig.~\ref{Almudallal_fig09_c}, using both methods.  As can be seen, for both the strong and the weak coupling cases, the differences between the $MH$ hysteresis loops calculated from the MEP (exact) formulation and direct path approximation are generally small and only weakly dependent on the sweep rate $R$.

\begin{figure}[ht]
\subfigure{
\centering\includegraphics[clip=true, trim=0 0 0 0, height=5.0cm]{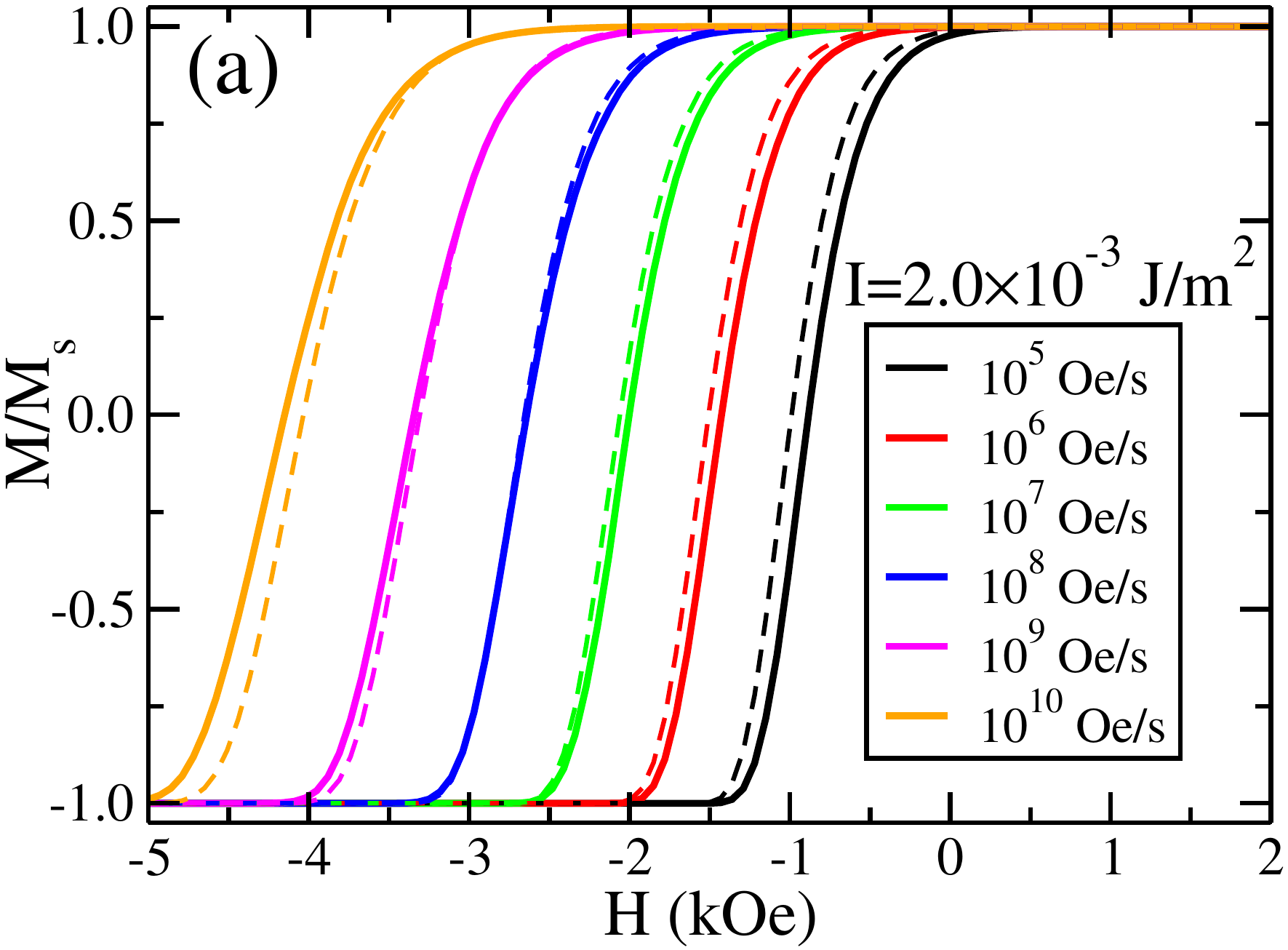}
\label{Almudallal_fig09_a}
}
\subfigure{
\centering\includegraphics[clip=true, trim=0 0 0 0, height=5.0cm]{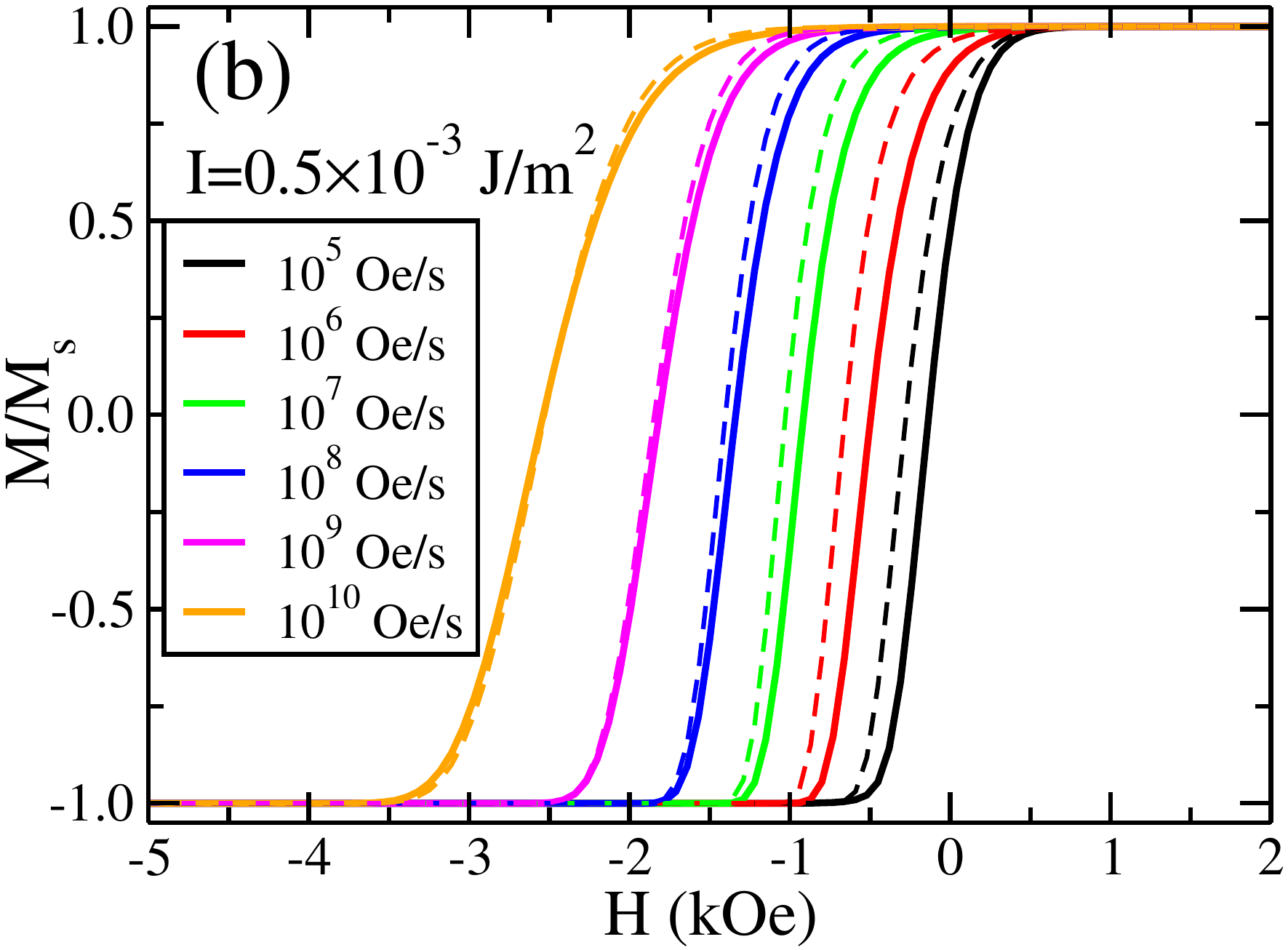}
\label{Almudallal_fig09_b}
}
\subfigure{
\centering\includegraphics[clip=true, trim=0 0 0 0, height=5.0cm]{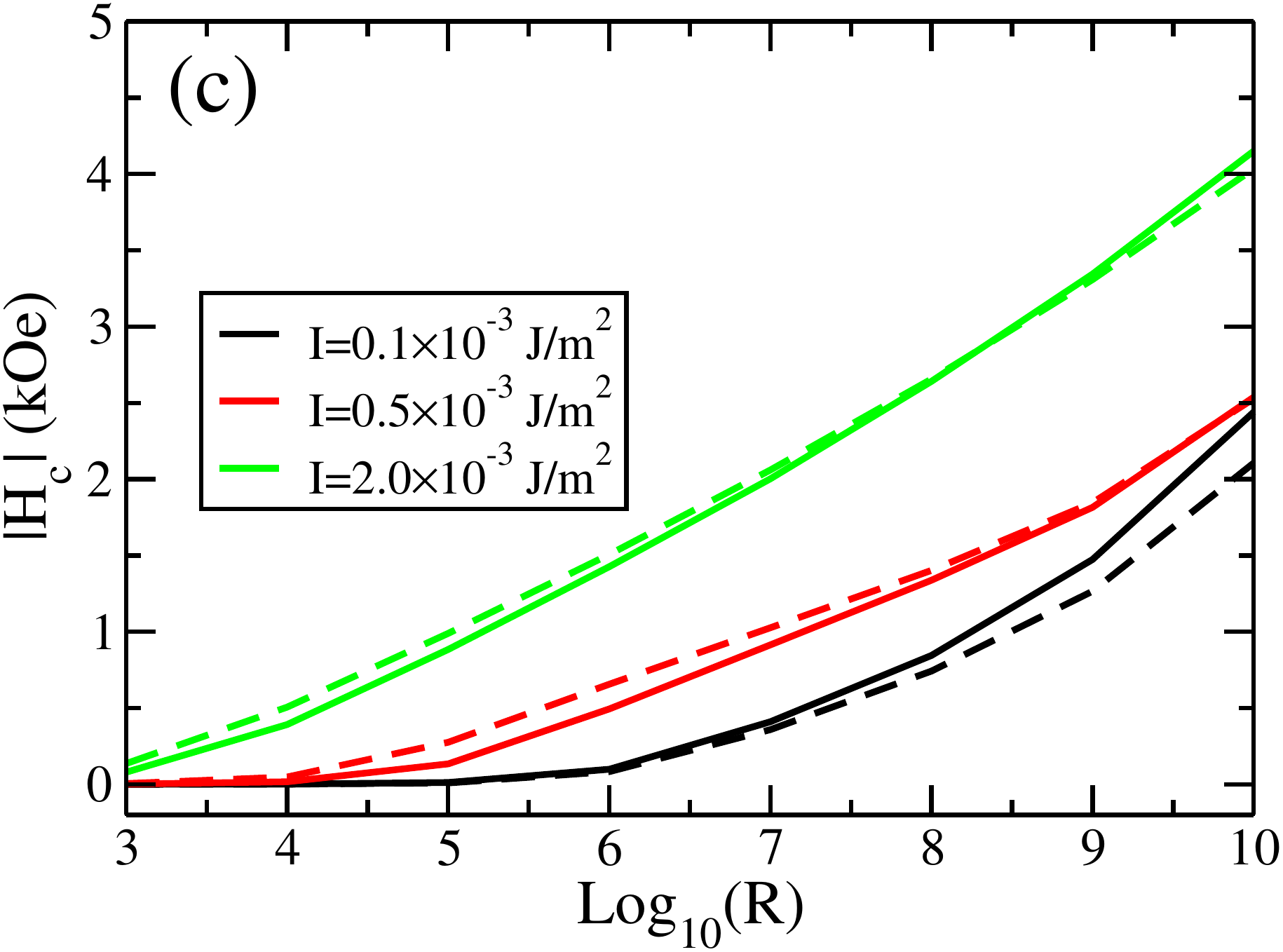}
\label{Almudallal_fig09_c}
}
\caption{(a) shows the $MH$ hysteresis loops at different sweep rates for $I$=2.0$\times 10^{-3}$\:J/m$^2$ and (b) for $I$=0.5$\times 10^{-3}$\:J/m$^2$. Solid lines are obtained from the MEP method and the dashed lines are from the direct path approximation. The extracted coercivity as a function of the sweep rate for $I$=2.0, 0.5, and 0.1$\times 10^{-3}$\:J/m$^2$ is shown in (c).} 
\label{Almudallal_fig09}
\end{figure}

One drawback of this approach is the fact that it is actually two distinct approximations, one valid for the strong coupling regime and another valid for the weak coupling regime, and it does not really provide an obvious way of interpolating between them.  

\subsection{Transient State Approximations and Metabasins}

The second approximation to consider is based on the fact that, depending on the parameters, there can be significant differences in the energy barriers and the attempt frequencies separating  the energy minima. By way of an example, the calculated energy barriers and attempt frequencies between minima are presented in Table \ref{table1} together with the calculated rate constants $r_{\alpha\beta}$ and the mean escape times $\tau_{\alpha\beta} = 1/r_{\alpha\beta}$ for the case $I = 0.5\;\times10^{-3}\:\mathrm{J/m^2}$ and $H=0$, shown in Fig.~\ref{Almudallal_fig02_a}. Because of the factor $\exp( -\Delta E/k_BT)$ in the Arrhenius-N\'{e}el expression, the differences in  $\Delta E_{\alpha\beta}$ (which are approximately $4 \thicksim 5$) can lead to rate coefficients that differ by several orders of magnitude. This suggests that some of the states, in this case specifically states $\sigma_2$ and $\sigma_3$, are very short lived and will not contribute significantly to the magnetization for processes involving long time scales (i.e. $MH$ hysteresis loops generated using the vibrating sample magnetometer (VSM)). However, one has to be careful in removing such transients as they serve as intermediate states in the process of grain reversal. 

\begin{table}
\begin{tabular}{| c | c | c | c | c |c | c | c | c | c |}
\hline
   & $\Delta E_{\alpha\beta}/k_BT$ & $f_{\alpha\beta}$ (GHz) & $r_{\alpha\beta}$ (MHz) & $\tau_{\alpha\beta}$ ($\mu$s) \\
\hline
$1\to 2$ & 12.4826 & 20.5832 & $7.80502\times 10^{-2}$  & $1.28121\times 10^{1}$  \\
$2\to 1$ & 3.79107 & 8.47542 & $1.91303\times 10^{2}$  & $5.22731\times 10^{-3}$  \\ \hline
$1\to 3$ & 19.6502 & 51.7315 & $1.51275\times 10^{-4}$ & $6.56705\times 10^{3}$\\
$3\to 1$ & 10.9587 & 21.3012 & $3.7078\times 10^{-1}$ & $2.69701$\\ \hline
$2\to 4$ & 10.9846 &  8.52533 & $3.60141\times 10^{-1}$ & $2.77668$\\
$4\to 2$ & 19.6762 &  51.5684 & $1.46935\times 10^{-4}$ & $6.75972\times 10^{3}$\\ \hline
$3\to 4$ & 3.78935 & 8.52533 & $1.92759\times 10^{2}$  & $5.18781\times 10^{-3}$\\
$4\to 3$ & 12.4809 & 20.7044 & $7.86445\times 10^{-2}$  & $1.27153\times 10^{1}$\\
\hline 
\end{tabular}
\caption{The energy barriers, attempt frequencies, rate coefficients and mean escape times calculated from the MEPs connecting the minimum energy states for the case $I = 0.5\;\times10^{-3}\:\mathrm{J/m^2}$ and $H=0$ corresponding to the energy landscape shown in Fig.~\ref{Almudallal_fig02_a}.}
\label{table1}
\end{table}

Consider for example state $\sigma_1$ with both grains aligned along the positive $z$-axis. It can make the transition to states $\sigma_2$ or $\sigma_3$. Comparing the mean escape times associated with the two transitions it is obvious, since $\tau_{1\to2}  \ll \tau_{1\to3} $, that the predominant transition will be to state $\sigma_2$. From state $\sigma_2$ the grain again has two choices. It can make the transition to state $\sigma_4$ or back to state $\sigma_1$. Comparing the mean escape times it is clear, since $\tau_{2\to1}  \ll \tau_{2\to4} $, that the predominant transition is for the grain to return to its initial state $\sigma_1$. This implies that grains in the states $\sigma_1$ and $\sigma_2$ will  fluctuate back and forth with a characteristic time scale of the order of $10\:\mu\mathrm{s}$ for some time before it will transition to $ 1 \to 3$ or $2\to 4$. The effect of these fluctuations will be to establish a local thermodynamic equilibrium between the two states $\sigma_1$ and $\sigma_2$ with a time scale on the order of a fraction of a ms. When local equilibrium is established, the net average probability flux between the two states will be zero and hence $\mathcal{I}_{1\to2} = \mathcal{I}_{2\to1}$. A similar argument may be applied to the states $\sigma_3$ and $\sigma_4$. 

The above argument implies that, while $p_1(t)$ and $p_2(t)$ are time dependant, they will nevertheless satisfy  the constraint,
\begin{align}
\frac{p_1(t)}{p_2(t)} = \frac{r_{21}}{r_{12}}=\exp\left( -\frac{\Delta G_{12}}{k_BT} \right),
\label{boltzmannRatio}
\end{align}
where  $\Delta G_{12} = G_{1} - G_{2}$  and $G_\alpha$ is expressed in terms of $\mathcal{Z}_\alpha$, defined in Eq.~\eqref{localEq}, as  
\begin{align}
G_\alpha = -k_BT \log \mathcal{Z}_\alpha.
\end{align}
This is consistent with the requirement that states in the metabasin $\Omega_A = \Omega_1 \cup \Omega_2$ satisfy the condition of local equilibrium $c_1(t) = c_2(t) = c_A(t)$ and hence the probability density within the metabasin formed by the union $\Omega_A = \Omega_1 \cup \Omega_2$ is given by a Boltzmann distribution $\rho_A(x,t) = c_A(t)\exp(-E(x,t)/k_BT)$. Again a similar argument can be made for grains in the states $\sigma_3$ and $\sigma_4$.

The above reasoning implies that for processes with time scales on the order of ms or greater, we can assume that $p_1(t)/p_2(t) = r_{21}/r_{12}$ and $p_4(t)/p_3(t) = r_{34}/r_{43}$. If we therefore define metabasins as those regions of phase space $\Omega_A = \Omega_1 \cup \Omega_2$ and $\Omega_B = \Omega_3 \cup \Omega_4$, then the probability of finding a grain in one of these metabasins is simply given by $p_A(t) = p_1(t) +p_2(t)$ and $p_B(t) = p_3(t) +p_4(t)$ which can be shown to satisfy the following rate equations, 
\begin{eqnarray}
\label{rate_weak_2state}
\frac{dp_A}{dt} &=& -r_{AB} p_A + r_{BA} p_B, \nonumber \\
\frac{dp_B}{dt} &=& -r_{BA} p_B + r_{AB} p_A, 
\end{eqnarray}
where the rate coefficients $r_{AB}$ and $r_{BA}$ are given by,  
\begin{align*}
r_{AB}=\frac{r_{13} r_{21} + r_{12} r_{24}}{r_{12} + r_{21}}, && r_{BA}=\frac{r_{31} r_{43} + r_{34} r_{42}}{r_{34} + r_{43}}. 
\label{metaRates}
\end{align*}
Using this concept of metastates, the set of four rate equations has been reduced to two, where $p_A(H)$ and $p_B(H)$ can be obtained by numerical integration.  The probabilities $p_\alpha(H)$ for $\alpha \in \{1,2,3,4\}$ can be determined from the values of $p_A(H)$ and $p_B(H)$ together with the ratios $r_{21}(H)/r_{12}(H)$ and $r_{43}(H)/r_{34}(H)$ and hence the magnetization calculated as a function of $H$. 

Fig.~\ref{Almudallal_fig10} shows a comparison of the $MH$ hysteresis loops for $I$=0.5$\times 10^{-3}$\:J/m$^2$ between the original four-state model (solid curves) and the two-state approximation (open circles). The two models show very good agreement up to $R \le 10^8\:\mathrm{Oe/s}$ above which the assumption of a Boltzmann probability distribution within the metastates $A$ and $B$ is no longer accurate. 
\begin{figure}[ht]
\centering\includegraphics[clip=true, trim=0 0 0 0, height=5.0cm]{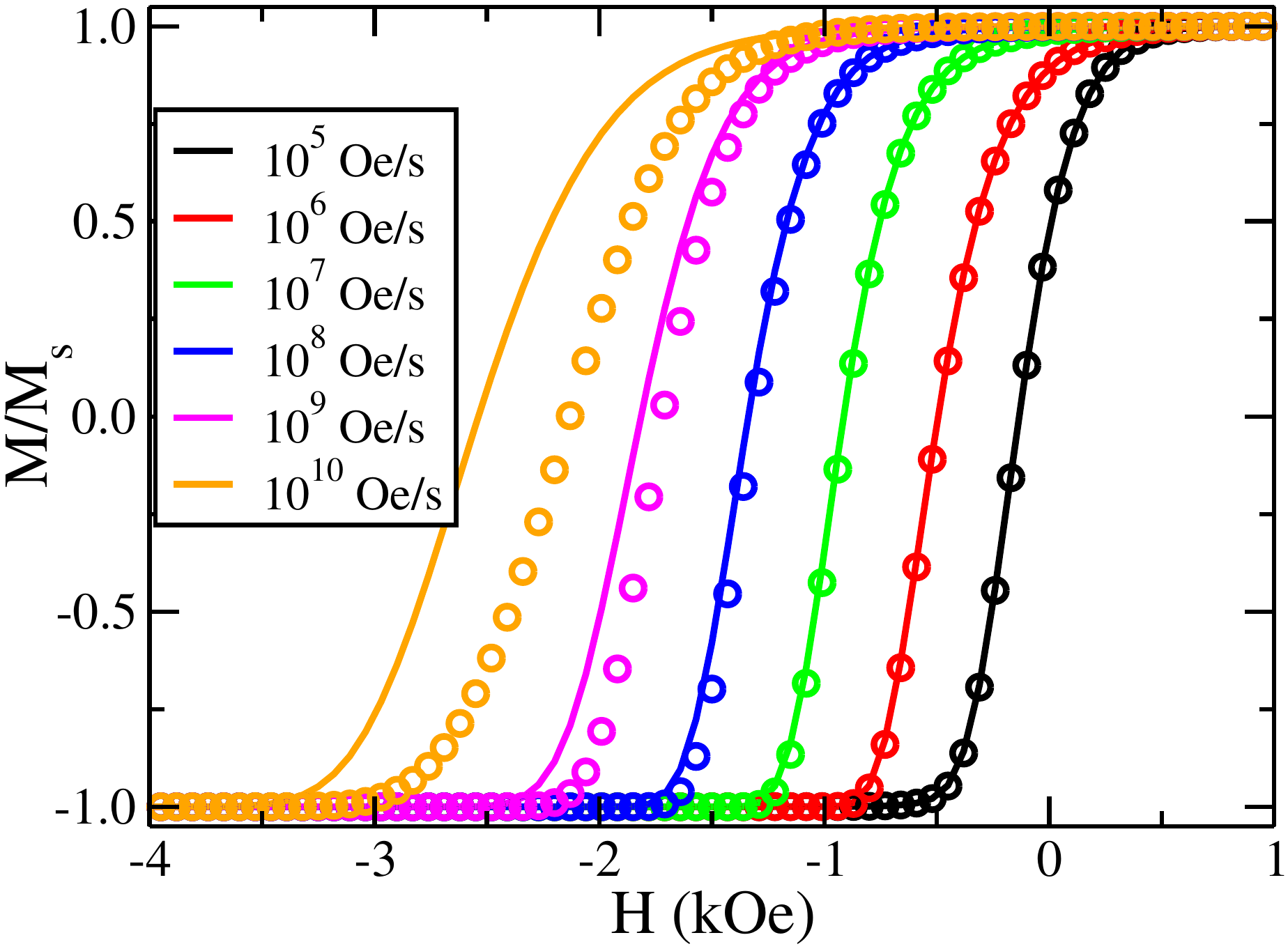}
\caption{A comparison of the $MH$ hysteresis loops for $I$=0.5$\times 10^{-3}$\:J/m$^2$ for the four-state model (solid lines) and the two-state approximation (open circles).}
\label{Almudallal_fig10}
\end{figure}
The above analysis in terms of metastates not only simplifies the system of equations that need to be solved for a range of $R$ values but also provides some insight into how to understand the complex relationship between the sweep rate $R$ and the response of ECC media. It is also important to note that while the case in which the system is described in terms of two metabasins, how the phase space up is divided into metastates for a given set of parameters is dependent on the nature of the energy landscape and time scales of interest. Indeed it is possible to adjust the number and regions of phase space occupied by the metabasins as the system evolves. In contrast to the previous approximation schemes, note that the two state model described in this section evolves  smoothly into the coherent rotation of the strong coupling case as the exchange coupling constant $I$ increases. 

\section{Metastates and Kinetic Monte Carlo}
\label{metastatesKMC}
When the present model is extended to include magnetostatic and intralayer exchange interactions, the direct integration of the rate equations is no longer feasible. An alternative approach is the KMC algorithm, which utilizes a stochastic algorithm to integrate the rate equations, and which can be adapted to include the interactions between the grains.\cite{mepkmc} However, the presence of low energy barriers can significantly increase the simulation time, effectively rendering the KMC approach no longer feasible at low sweep rates. This is a longstanding problem with KMC simulations.\cite{voter2007} One way of dealing with this problem is by combining clusters of minimum energy states separated by low energy barriers into ``metabasins" as described in the previous section. 

To demonstrate the significance of the role of  metabasins in the application of the KMC algorithm, consider the decay of an initially fully polarized ensemble of $N$ identical non-interacting grains ($p_1=1$) with zero field and $I=0.5\times10^{-3}\:\mathrm{J/m^2} $. Using the rate coefficients presented in Table \ref{table1}, the wait times for each of the $N$ grains is given by
\begin{align}
t_{\alpha\to\beta} (n) = r^{-1}_{\alpha\beta} \log(x),
\end{align}
where $x$ is a uniformly distributed random number $\in \{0 < x < 1\}$, $\alpha$ denotes the state of the $n^\mathrm{th}$ grain and $\beta$ represents the two possible states it can transition into. The wait times describe how long we might expect to wait before the $n^\mathrm{th}$ grain would make the transition $\alpha \to \beta$. The shortest of these wait times defines the first reversal time. The KMC step then takes the transition with the shortest reversal time and switches the grain from state $\alpha$ to a new state $\beta$. This process is then repeated generating a stochastic sequence that models the process of thermally activated grain reversal.   
\begin{figure}[ht]
\centering\includegraphics[clip=true, trim=0 0 0 0, height=5.0cm]{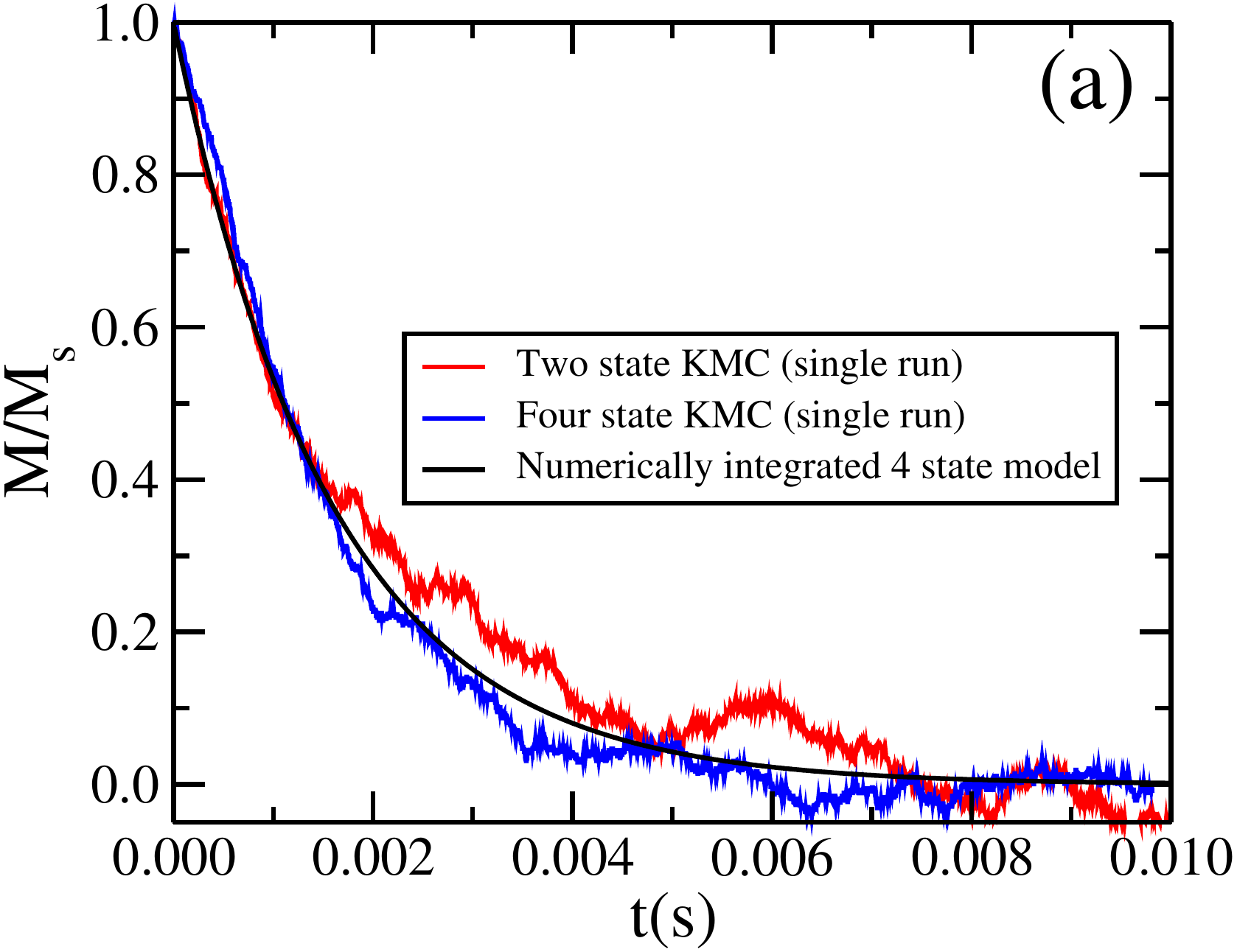}
\centering\includegraphics[clip=true, trim=0 0 0 0, height=5.0cm]{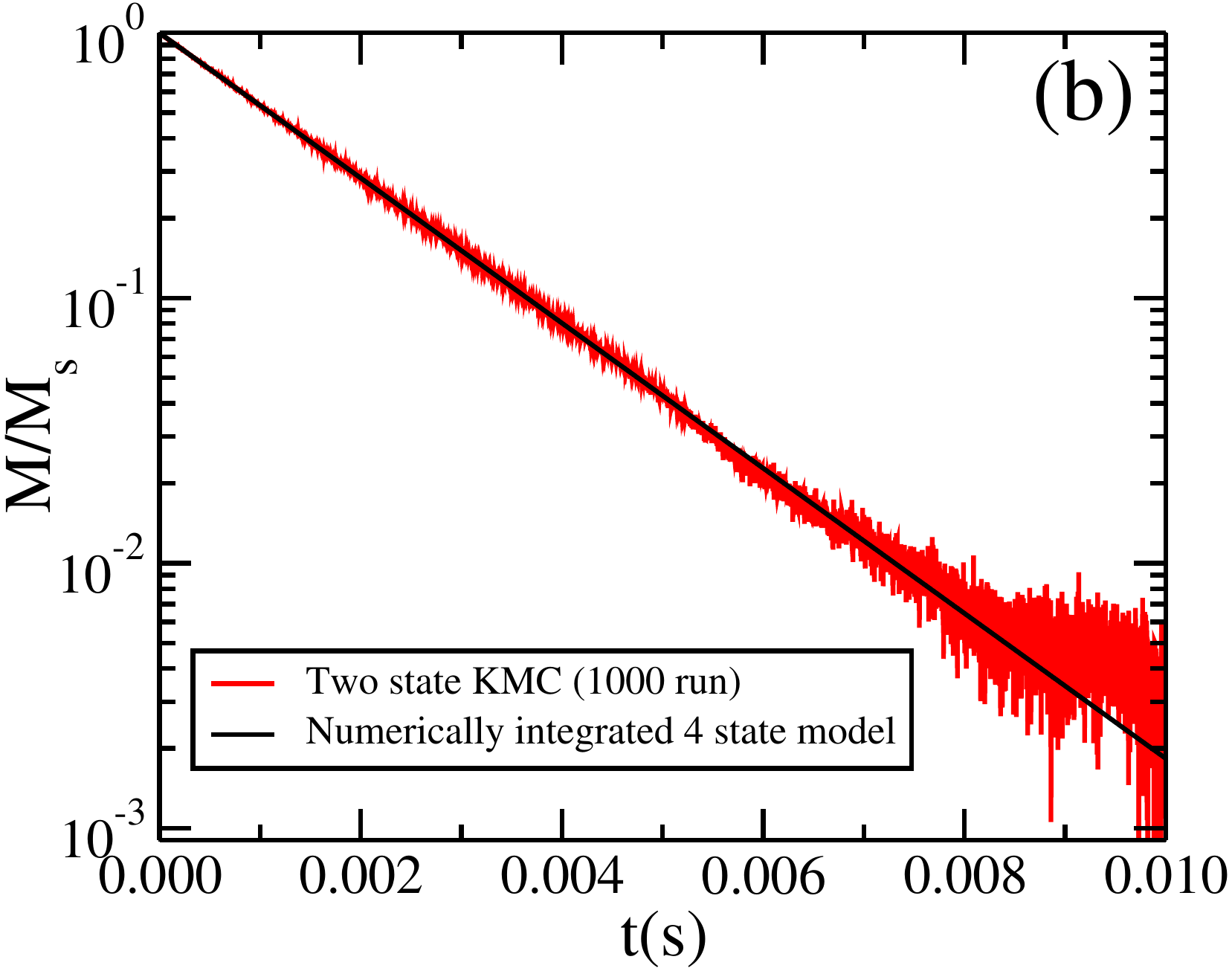}
\caption{ Decay of the normalized magnetization at zero applied field with $I$=0.5$\times 10^{-3}\:\mathrm{J/m^2}$ (a) for the two-state model (red line) and the four-state model (blue line) from the KMC method together with the numerical integration of the four state model (black line) and (b) comparison of KMC results for two state model averaged over 1000 runs (red line) together with results obtained from numerical integration of the four state model (black line).}
\label{Almudallal_fig11}
\end{figure}

Figure \ref{Almudallal_fig11} (a) shows the magnetization $m$ plotted as a function of $t$ calculated using the KMC method for both the four state model and the two ``metastate" representation for a system of 1000 non-interacting grains, together with the numerical solution of the rate equations for the four state model. The KMC solutions show the effects of the stochastic fluctuations and both are in reasonable agreement with the solution obtained by direct integration of the rate equations.  However, for the four state model, the average first reversal time was $6.1038\times 10^{-9}\:\mathrm{s}$ while for the two metastate representation, the average first reversal time was $3.1595\times 10^{-6}\:\mathrm{s}$; a factor of approximately $500$ times greater than the four state case. This difference arises from the fact that vast majority of KMC steps in four state model were simply fluctuations within the metabasins A ($\sigma_1 \leftrightarrow \sigma_2$) and B ($\sigma_3 \leftrightarrow \sigma_4$). The difference in the average KMC time step is reflected in the run times; 39 minutes in the case of the four state model and approximately 4 seconds in the case of the two metastate representation. The speed up factor of 600 in completion times for the four and two state representations includes not only the shorter time steps but also the computational overhead associated with the four state model. To demonstrate the equivalence of the results obtained from the two state KMC calculations and those obtained by the direct integration of the four state model, Fig.~\ref{Almudallal_fig11} (b) shows good agreement between a plot of the average $m\,\mathrm{vs.}\, t$ obtained from the two metastate representation averaged over 1000 independent KMC runs and those obtained by direct integration of the four state model. 

These results illustrate that for future applications with interacting grains, where direct integration of the rate equations is not feasible and the time scales of stochastic LLG restricts its application to $\mu\mathrm{s}$ time scales, the KMC approach represents a viable model of long-time processes dominated by thermally activated reversal. Further, when the system in question, such as ECC media, has a range of energy barriers, the above example demonstrates that removing the short time scale fluctuations associated with transient states by combining them into a single metabasin can result in significant computational efficiencies with negligible loss of accuracy. In the case of interacting systems, the gains in run time are even more significant given the increased computational overhead involved in computing the effective fields due to the interactions and the more complex energy landscapes that typically include a greater number of critical points than the simple model discussed here.  

\section{Discussion and Conclusions}
\label{conclusions}

A set of rate equations are presented that describe the evolution of a non-interacting ensemble of dual layer ECC grains based on processes of thermally activated grain reversal. The rate coefficients are calculated from the Langer formalism and have the Arrhenius-N\'{e}el form in which the attempt frequency and energy barriers are expressed in terms of the energy and its Hessian matrix calculated at the maximum point on the minimum energy paths that connect the energy minima. The particular form for the attempt frequency is outlined in the Appendix and is not restricted to the canonical coordinates commonly used in the derivation but is valid for any system (or systems) of generalized coordinates that parameterize the surface of a sphere. The minimum energy paths are calculated using the so called ``string method''. The rate equations can be integrated numerically for the case of a time dependent applied field with a constant sweep rate and the magnetization calculated to produce $MH$ hysteresis loops.

It is shown that the method may be used to study both the strong coupling regime, in which the energy landscape has two energy minima, consisting of two ferromagnetically aligned layers as well as the more complicated weak coupling regime, which has an energy landscape that can have up to four distinct energy minima, two ferromagnetic and two antiferromagnetic states. Calculating the $MH$ hysteresis loops therefore requires solving two coupled rate equations for the strong coupling regime and up to four coupled rate equations for the weak coupling regime. The results show that, for the parameters used in the current work, the transition from the weak to the strong coupling regime occurs when $I \sim $1.0$-$1.5$\times 10^{-3}$\:J/m$^2$, which is the region of interest for ECC based recording media.  

Verification of our model results for $MH$ hysteresis loop was achieved through comparison with LLG simulations on a dual layer system, each layer with a $16\times16$ non-interacting grains.  The high degree of agreement confirms the accuracy of the rate coefficients and the numerical integration of the rate equations.  In addition, using the $MH$ hysteresis loops calculated from the rate equations, the effect of rate dependence and exchange coupling on a Figure of Merit based on the ratio between the energy barrier and switching field was calculated. This provides some guidance on the optimal coupling between the layers. 

Results based on a direct path approximation to the MEP in the strong and weak coupling limits that permit analytic expressions for both the energy barriers and the attempt frequencies are presented in Figs.~\ref{Almudallal_fig09}(a) and \ref{Almudallal_fig09}(b). The results  show remarkably good agreement with the exact MEP calculation for both the strong, $I =2.0\times10^{-3}\:\mathrm{J/m^2} $, and the weak, $I =0.5\times10^{-3}\:\mathrm{J/m^2}$, coupling regimes.

Another approximation scheme was presented in which pairs of minima separated by a relatively low energy barrier so that they are very rapidly equilibrated and could be combined into a single metastate in which the ratio $p_\alpha/p_\beta$ is given by a Boltzmann factor (Eq.~\eqref{boltzmannRatio}). It was shown that for the case $I =0.5\times10^{-3}\:\mathrm{J/m^2}$, the $MH$ hysteresis loops obtained by integrating the four state rate equations could be accurately reproduced by integrating the rate equations for a two ``metastate" representation with rate coefficients between the metastates given by Eq.~\eqref{rate_weak_2state}. The potential importance of this mapping of ``exact" the four state model to a model consisting of two metastates was demonstrated in simulation of magnetic decay using the KMC algorithm, in which the two ``metastate" model produced results essentially equivalent to the four state model, but with a run time that was reduced by a factor of 600. 

The results of this work serve as a prelude to the extension of our previous KMC approach\cite{fal1} to study thermally activated magnetic grain reversal in dual layer ECC media that includes magnetostatic and intralayer exchange interactions.\cite{mepkmc} 
The essentially exact treatment of grain reversal for the dual layer ECC grain problem as outlined in this work, serves as the foundation for this extension, while combining cluster of states that are separated by  relatively small energy barriers to form metastates allows us to deal with the phenomena of ``stagnation" that can severely limit the accessible run times that can be achieved using the KMC approach.

This extension of our previous KMC algorithm will allow for the direct comparison of experimentally determined slow-sweep-rate $MH$ hysteresis loops for ECC media with corresponding modelled results. This capability is useful for the estimation of model parameters which characterize recording media such as intralayer and interlayer exchange couplings.  Such a direct comparison is not possible with traditional LLG simulations where long time scales are inaccessible. This dual layer KMC algorithm will also be especially useful in applications to dual layer media for heat assisted magnetic recording where thermally activated moment reversal is pronounced.\cite{plumer4} In addition, the investigation of magnetostatic and intralayer interaction effects on the FOM of Fig.~\ref{Almudallal_fig08} is of particular interest. 

\section{Acknowledgments}

This work was supported by Western Digital Corporation, the Natural Science and Engineering Research Council (NSERC) of Canada, the Canada Foundation for Innovation (CFI), and the Atlantic Computational Excellence network (ACEnet). We thank I. Saika-Voivod for numerous insightful discussions. 

\appendix

\section{}
The attempt frequency given in Eq.~\eqref{attemptFrequency} is key to the analysis presented in the previous sections, we therefore outline the derivation in some detail. The approach starts with the Fokker-Planck equation (FPE) for a single magnetic moment in a magnetic field and the generalization to consider a set of exchanged coupled moments. 

The FPE can be derived from the stochastic LLG equation.\cite{brown79,gardiner,langer} From the FPE, the rate constants $r_{\alpha\beta}$ defined by Eq.~\eqref{ArrheniusNeel} are calculated using the formalism presented by Langer\cite{langer} adapted to account for the dissipative dynamics of magnetic moment  in an applied field.\cite{Coffey2001} While the application of the Langer formalism is facilitated by a judicious choice of coordinates, $(\phi,z=\cos\theta)$, often referred to as the canonical coordinates, it is nevertheless possible derive a straightforward expression for the rate coefficients based on any set of generalized coordinates $(u^1,u^2)$ that parameterize the surface of the sphere $\mathbb{S}$ using as basis vectors the covariant tangent vectors $\vec g_i \equiv \partial \hat m/\partial u^i$. An advantage of this approach is that it allows the direct application of the tools of differential geometry to be applied to the problem. This is of some practical importance in the case of spin dynamics as it is not possible to define a single coordinate system on the surface of a sphere where the metric is everywhere finite. However, the surface of a sphere can be treated as a differentiable manifold by dividing it into overlapping regions, each of which has a metric that is everywhere finite.       

Consider a magnetic moment $\vec m_a$ of volume $v_a$ with magnetization $M_a$, anisotropy constant $K_a$, and a damping factor $\alpha_0$ in a magnetic field $\vec H$. The equation of motion for the moment is given by
\begin{align}
\frac{d\hat m_a}{dt} = \gamma \mu_0\left( \hat m_a\times  \vec H - \frac{\alpha_0}{1+\alpha_0^2}\hat m_a \times \left( \hat m_a \times\vec H\right)\right),
\end{align}
where $\hat m_a =\vec m_a/M_a v_a$ and $\vec H = -\mu_0^{-1} \partial E/\partial \vec m_\alpha$. We parameterize the unit vector $\hat m_a$ in terms of the generalized coordinates $u=(u^1,u^2)$ (e.g. $u=(\theta_a,\phi_a)$) which cover the surface of the unit sphere. Since $d\hat m_a/dt$ will be tangential to the surface of the sphere $\mathbb{S}_a$, we define the local covariant basis vectors\cite{kreyszig}
\begin{align}
\vec g_i = \frac{\partial \hat m_a}{\partial u^i}. 
\end{align} 
Any vector tangential to the surface of the sphere can therefore be written as $\vec v = \vec g_i v^i$, where the components $v^i$ define a type (1,0) tensor. The basis vectors $\vec g_i$ are, in general, neither orthogonal nor normalized to unity, but satisfy
\begin{align}
\vec g_i \cdot \vec g_j = \bar g_{ij},
\end{align}
where $\bar g_{ij}$ is the metric tensor. We also define the reciprocal, or contravariant, basis vectors  $\vec g^i$ such that 
\begin{align}
\vec g_i \cdot \vec g^{\,j} = \delta_i^j,
\end{align}
where $\delta_i^j$ is the Kronecker delta function. Any tangential vector $\vec v$ may then also be written in contravariant form as
\begin{align}
\vec v = \vec g^{\,i} v_i.
\end{align}
The scalar product of any two tangential vectors $\vec u$ and $\vec v$ may then be written as 
\begin{align}
\vec u \cdot \vec v = u_iv^i = \bar g^{ij} u_i v_j = \bar g_{ij}u^iv^j,
\end{align}
where the components $v_i$ define a type (0,1) tensor. The vector $\hat m_a$ may also be written in terms of the basis vectors $\vec g_i$ and $\vec g^{\,i}$ as  
\begin{align}
\hat m_a = \frac{\vec g_1 \times \vec g_2}{\left|\vec g_1 \times \vec g_2 \right|}= \frac{\vec g_1 \times \vec g_2}{\left|\vec g^1 \times \vec g^2 \right|}.
\end{align}
It is straightforward to show that
\begin{align}
\hat m_a \times \vec g_i &= \sqrt{\bar g}\epsilon_{ij}\,\vec g^{\,j}\\
\hat m_a \times \vec g^{\,i} &= \frac{\epsilon^{ij}}{\sqrt{\bar g}}\,\vec g_j,
\end{align}
where  $\epsilon_{ij}$ and $\epsilon^{ij}$ denote the Levi-Cevita symbols defined as
\begin{align}
\epsilon_{11} = \epsilon_{22} =0, \quad \epsilon_{12} = -\epsilon_{21} = 1,\quad \epsilon^{11} = \epsilon^{22} = 0,  \quad \epsilon^{12} = -\epsilon^{21} = 1,
\end{align}
and $\bar g=\hat m_a\cdot(\vec g^1\times\vec g^2)$, which is simply the volume of the vectors triad $\left(\hat m_a, \vec g_1,\vec g_2\right)$. It can be shown that $\epsilon^{ij}/\sqrt{\bar{g}}$ and $\sqrt{\bar{g}}\epsilon_{ij}$ define tensors of the form (2,0) and (0,2), respectively, on the surface $\mathbb{S}$. The LLG equation can then be written in covariant form as 
\begin{align}
v^i &=\frac{du^i}{dt}=-\gamma_B\left( \frac {\epsilon^{ij}}{\sqrt{\bar g} }  - \frac{\alpha_0}{1+\alpha_0^2}\bar g^{ij}\right)\mu_0 H_j\nonumber \\ 
&=\frac{\gamma_B}{m_a}\left( \frac {\epsilon^{ij}}{\sqrt{\bar g} }  -  \frac{\alpha_0}{1+\alpha_0^2}\bar g^{ij}\right)\frac{\partial E(u) }{\partial u_j} \label{velocity}.
\end{align}
We note that the form of the above equations are invariant under a generalized coordinate transformation.  

Consider an ensemble of such spins and denote by $\rho(u,t)$ the probability density, then the probability of a single spin will be aligned in the solid angle $d\Omega$ at time $t$ is given by $dp(u,t) = \rho(u,t) d\Omega$. The probability density $\rho(u,t)$ may be calculated from the Fokker-Planck equation (FPE) which can be written in terms of the coordinates $u^i$ as 
\begin{align}
\frac{\partial \rho(u,t)}{\partial t} &=- \nabla_i J^i(u,t), 
\end{align}
where the probability current density $J^i\left( u,t\right)$ consists of an advective term and a diffusive term
 \begin{align}
J^i\left( u,t\right) &= \rho(u,t) v^i -\frac{\gamma_B^2D_a }{1-\alpha_0^2}\bar g^{ij} \nabla_j \rho(u,t),
\end{align}
where $D_a= \alpha_0 k_BT/\gamma_Bm_a$, with $m_a=M_av_a$, the velocity field $v^i$ is given by Eq.~\eqref{velocity} and $\nabla_i$ denotes the absolute derivative, and 
\begin{align}
\nabla_i \rho(u) &= \frac{\partial \rho(u)}{\partial u_i}\\
\nabla_i v^i(u) &= \frac{1}{\sqrt{\bar g}}\frac{ \partial \left(\sqrt{\bar g}v^i(u)\right)}{\partial u_i} =\frac{\partial v^i(u)}{\partial u^i} + v^i(u)\Gamma_{ij}{}^j,
\end{align}
where $\Gamma_{ij}{}^k$ denotes the Christoffel symbol of the second kind.\cite{kreyszig} Since we are interested in solutions close to equilibrium, following Langer,\cite{langer} we write the probability density in terms of the crossover function $c(u,t)$ as
\begin{align}
\rho(u,t) = c(u,t) \exp\left(- \frac{E_a(u,t)}{k_B T}\right).
\end{align}
It can then be shown that $J^i(u,t)$ may be written in terms of the crossover function $c(u,t)$ as 
\begin{align}
J^i(u,t) = &k_BT\frac{\gamma_B}{m_a}\exp\left(-\frac{E_a(u,t)}{k_BT}\right)\left( \frac{\epsilon^{ij} }{\sqrt{\bar g} }- \frac{\alpha_0}{1- \alpha_0^2} \bar g^{ij} \right) \nabla_j c(u,t)\nonumber \\
& \quad + \text{divergenceless terms}.
\end{align}
The above formalism can be readily extended to the problem of two coupled spins. Let $w =(w^1,w^2)$ denote the generalized coordinates that  specify the orientation of a second magnetic moment of volume $v_b$, magnetization $M_b$, anisotropy constant $K_b$ and damping constant $\alpha_0$. In the absence of the interaction, the probability current density on the surface of the sphere $\mathbb{S}_b$ may then be written as 
\begin{align}
J^i(w,t) = &k_BT\frac{\gamma_B}{m_b}\exp\left(-\frac{E_b(w,t)}{k_BT}\right)\left( \frac{\epsilon^{ij} }{\sqrt{\bar g} }- \frac{\alpha_0}{1+ \alpha_0^2} \bar g^{ij} \right) \nabla_j c(w,t)\nonumber \\
& \quad + \text{divergenceless terms}.
\end{align}

In presence of an interaction between the moments, we define the vectors $x = (u,w)$ that spans the tangent space of the four dimensional manifold $\mathbb{S} = \mathbb{S}_a\otimes\mathbb{S}_b$. The metric $g^{\mu\nu}$ can be written in matrix form as 
\begin{align}
||g^{\mu\nu}|| = \left(\begin{matrix} ||\bar g^{ij}(a)|| & 0 \\
0 & ||\bar g^{ij}(b)||
\end{matrix}\right),
\end{align}
where $||\bar g^{ij}(a)||$ and $||\bar g^{ij}(b)||$ denote the matrix forms for the metrics in the manifolds $\mathbb{S}_a$ and $\mathbb{S}_b$ for the single grains $a$ and $b$. The LLG equation for the case of interacting spins may then be written in covariant form as 
\begin{align}
\frac{dx^\mu}{dt}&=-\frac{\gamma_B}{m}T^{\mu\nu}(x)\frac{\partial E(x,t) }{\partial x^\nu}, \label{velocity3}
\end{align}
where $m=m_a+m_b$ and the tensor $T^{\mu\nu}(x)$ expressed in matrix form as 
\begin{align}
||T^{\mu\nu}|| = \left(\begin{matrix} 
 \dfrac{m}{m_a}\left(\dfrac{\alpha_0}{1+\alpha_0^2}||\bar{g}^{ij}(a)|| - \dfrac{||\epsilon^{ij}||}{\sqrt{g(a)}}\right) &0\\
0 & \dfrac{m}{m_b} \left(\dfrac{\alpha_0}{1+\alpha_0^2}||\bar{g}^{ij}(b)|| - \dfrac{||\epsilon^{ij}|| }{\sqrt{g(b)}}\right)
\end{matrix}\right).
\end{align}

This yields the following expression for the probability current density in terms of the crossover function $c(x,t)$
\begin{align}
J^\mu(x,t) = & - k_BT\left(\frac{\gamma_B}{m}\right)\exp\left(-\frac{E(x,t)}{k_BT}\right)T^{\mu\nu}(x) \nabla_\nu c(x,t) \nonumber \\
& \quad + \text{divergenceless terms}.
\label{Jmu1}
\end{align}
This gives 
\begin{align}
\nabla_\mu J^\mu(x,t) = -k_BT \frac{\gamma_B}{m} \exp\left(-\frac{E}{k_BT}\right) \left(\dfrac{\alpha_0}{1+\alpha_0^2} \nabla_\mu G^{\mu\nu} - \dfrac{1}{k_BT} \dfrac{\partial E}{\partial x_\mu}T^{\mu\nu}\right)\nabla_\nu c(x,t),
\end{align}
where $G^{\mu\nu}$ may be written in matrix form as  
\begin{align}
\left|\left| G^{\mu\nu}\right|\right|= \left(\begin{matrix} 
 \dfrac{m}{m_a}||\bar{g}^{ij}(a)|| &0\\
0 & \dfrac{m}{m_b} ||\bar{g}^{ij}(b)||
\label{divJ}
\end{matrix}\right).
\end{align}

As discussed in Secs. \ref{strongExchange}  and \ref{weakExchange} we are interested in stationary solutions that satisfy $\nabla_\mu J^\mu(x) = 0$ for which the crossover function is essentially homogeneous except in a narrow region in the neighbourhood of the boundaries $\Gamma_{\alpha\beta}$ where it goes from $c_\alpha\to c_\beta$ on crossing the boundary from $\Omega_\alpha \to \Omega_\beta$. These solutions correspond to a state of ``local'' equilibrium with thermodynamic equilibrium corresponding to the special case $c_\alpha = \mathrm{const}$ for all $\alpha$. In addition, as discussed in Sec. \ref{energyLandscapes}, for the energy scales we are interested in, the probability current density is concentrated in a narrow region surrounding the saddle point $s_{\alpha\beta}$ on the boundary $\Gamma_{\alpha\beta}$. The crossover function is thus required only in region surrounding  $s_{\alpha\beta}$. This allows for two approximations that simplify Eq. \eqref{Jmu1}. The first is to assume that the coordinate system is chosen such that the metric $g^{\mu\nu}$ does not have any singularities close to the saddle point and it can be approximated as a constant. The second, assumes a quadratic approximation for the energy
\begin{align}
E(x) &\approx E(x_s) +\frac12\sum \left.\frac{\partial^2 E(x)}{\partial x^\mu\partial x^\nu}\right|_{x=x_s}(x-x_s)^\mu(x-x_s)^\nu+ \dots\label{quadraticE}
\end{align}
Defining the eigenvectors  and eigenvalues of the Hessian matrix $\partial^2 E(x)/\partial x^\mu\partial x^\nu|_{x=x_s}$
as 
\begin{align}
\frac{\partial^2 E(x)}{\partial x^\mu\partial x^\nu} a^\nu_n =\lambda_n a^\mu_n, 
\end{align} 
we define the new coordinates $y^n$ as 
\begin{align}
y^n = \bar a_\mu^n \left(x-x_s\right)^\mu,
\end{align}
with $\bar a^m_\nu a^\nu_n =\delta_{mn}$.  The quadratic form of the energy may then be written as  
\begin{align}
E(x) \approx E(x_s) + \frac12 \sum_{n=1} ^4 \lambda_n\left( y^n\right)^2+ \dots
\end{align}
Note that $\lambda_1>\lambda_2> 0$, $\lambda_3 =0$ (by symmetry), and $\lambda_4 <0$. From Eq. \eqref{divJ} we then obtain in the static limit ($\nabla_\mu J^\mu(x) = 0$) the following equation for the crossover function
\begin{align}
\sideset{}{'}{\sum}_{m,n} \left( \dfrac{\alpha_0}{1+\alpha_0^2} \frac{\partial}{\partial y^m}\tilde G^{mn} -\frac{1}{k_BT}  y^m \lambda_m \tilde T^{mn}\right) \frac{\partial c(y)}{dy^n}=0,
\end{align}
where $\tilde G^{mn} = a^m_\mu G^{\mu\nu} a ^n_\nu$, $\tilde T^{mn} = a^m_\mu T^{\mu\nu} a ^n_\nu$  and $\sum^{'}$omits the term $m =3$ and $n= 3$.  As discussed by Langer,\cite{langer} this equation may be solved using the method of characteristics, whereby we look for solutions of the form $c(y) =c(t)$ where the variable $t$ defines a trajectory $t = \sum_{n\ne 3} U_n y^n$, with the direction cosines $U_n$ are given by the solutions of the eigenvalue equation 
\begin{align}
 \sideset{}{'}{\sum}_{n} \lambda_m \tilde T^{mn} U_n = \xi U_m. \label{eigenvalue}
\end{align}
The solution of interest is given by
\begin{align}
\frac{dc(t)}{dt} = C_0 \exp\left(-\frac{|\kappa| t^2}2\right)
\end{align}
with $\kappa=\xi(1+\alpha_0^2)/\alpha_0 G_U k_BT$, where $\xi$ denotes the negative eigenvalue obtained form Eq.~\eqref{eigenvalue} with $G_U = G^{mn}U_mU_n$. Integrating this equation using the boundary conditions $\lim_{t \to -\infty}c(t) = p_\alpha/\mathcal{Z_\alpha}$ and $\lim_{t \to \infty}c(t) = p_\beta/\mathcal{Z_\beta}$ gives 
\begin{align}
\frac{dc(t)}{dt} =\sqrt{\frac{|\kappa|}{2\pi}} \left(\frac{p_\beta}{\mathcal{Z}_\beta} - \frac{p_\alpha}{\mathcal{Z}_\alpha} \right) \exp\left(-\frac{|\kappa| t^2}2 \right).
\end{align}

Writing $t$ in terms of the direction cosines $U_n$ gives $\partial c(y)/\partial y^n = U^n dc/dt$ leads to the following expression for the probability current density in the region around the critical point
\begin{align}
\tilde J^m(y) =& \frac{\alpha_0}{1 + \alpha_0^2} G_U\sqrt{\frac{|\kappa|^3}{2\pi}}\left(k_BT\right)^2  
\left(\frac{p_\beta}{\mathcal{Z}_\beta} - \frac{p_\alpha}{\mathcal{Z}_\alpha} \right)\exp\left(-\frac{E_s}{k_BT}\right)\nonumber\\&
\quad\frac{U_m}{\lambda_m}\exp\left(- \frac1{2}\sum_{nk}{}^{'}\left(\frac{\lambda_k}{k_BT}\delta_{nk} + |\kappa| U_nU_k \right)y_ny_k \right),\label{Jy0}
\end{align}
for $m \ne 3$ ($J^3 = 0$). 

To calculate the net probability current $\mathcal{I}_{\alpha \leftrightarrow\beta}$ flowing between the basins of attractions $\Omega_\alpha$ and $\Omega_\beta$, we simply integrate the $\tilde T^4$ component of probability current density over the hypersurface defined by $y_4 = 0$ to give\cite{lovelock} 
\begin{align}
\mathcal{I}_{\alpha \leftrightarrow\beta} &= \int_{\Gamma_{\alpha\beta}} \left.\sqrt{\tilde g_s(y)}J^4(y)\right|_{y^4=0} dy_1dy_2 dy_3,
\end{align}
where $\tilde g_s(y)$ is defined in terms of the metric associated with the subspace formed by the vectors $\{y^1,y^2,y^3\}$ 
\begin{align}
\tilde g_s(y)=\det\left[ \begin{matrix} \tilde g_{11}(y) &\tilde g_{12}(y)&\tilde g_{13}(y)\\
\tilde g_{21}(y) &\tilde g_{22}(y)&\tilde g_{23}(y)\\
\tilde g_{31}(y) &\tilde g_{32}(y)&\tilde g_{33}(y)
\end{matrix}\right].
\end{align}

Because of the exponential factor in the expression for the probability current density, Eq.~\eqref{Jy0}, only the region in the immediate vicinity of $y^1 = y^2 = 0$ will contribute to the integral and we can therefore use the quadratic form of the energy given by Eq.~\eqref{quadraticE}. Also it is convenient to choose $y_3$ so that it corresponds to the azimuthal angle $\Phi = (\phi_a + \phi_b)/2$ as the integration with respect to $y_3$ simply yields a factor of $2\pi$. The net probability current $\mathcal{I}_{\alpha  \leftrightarrow \beta}$ may then be evaluated to give
\begin{align}
\mathcal{I}_{\alpha \leftrightarrow \beta} =\sqrt{\tilde{g}(s)} \frac{\gamma_B}m  G_U \frac{\alpha_0}{1+\alpha_0^2}| \kappa| \sqrt{(2\pi k_BT)^3}{}\left(\frac{p_\beta}{\mathcal{Z}_\beta} - \frac{p_\alpha}{\mathcal{Z}_\alpha} \right)\sqrt{\frac{(2\pi k_BT)^3}{|\lambda_1\lambda_2\lambda_4|}}\exp\left(-\frac{E_s}{k_BT}\right).
\end{align}
Writing the net probability current as $\mathcal{I}_{\alpha \leftrightarrow\beta}= \mathcal{I}_{\beta \to \alpha}-\mathcal{I}_{\alpha \to \beta}$ yields
\begin{align}
\mathcal{I}_{\alpha \to \beta} =-\sqrt{\tilde g(s)} \frac{\gamma_B}m  G_U \frac{\alpha_0}{1+\alpha_0^2}| \kappa|\left(\frac{p_\alpha}{\mathcal{Z}_\alpha}\right)\sqrt{\frac{(2\pi k_BT)^3}{|\lambda_1\lambda_2\lambda_4|}}\exp\left(-\frac{E_s}{k_BT}\right),
\end{align}
and hence the following expression for the rate constants $r_{\alpha\beta}$
\begin{align}
r_{\alpha\beta}=\sqrt{\tilde{g}(s)} \frac{\gamma_B}m  G_U \frac{\alpha_0}{1+\alpha_0^2}| \kappa|\left(\frac{\exp\left({-E_s/k_BT}\right)}{\mathcal{Z}_\alpha}\right)\sqrt{\frac{(2\pi k_BT)^3}{|\lambda_1\lambda_2\lambda_4|}}.\label{rateCoefficient}
\end{align}

In order to compute $\mathcal{Z}_\alpha$, again assume that the probability density is strongly localized at $\sigma_\alpha$ and, since $\mathcal{Z}_\alpha$ is a scalar quantity, transform from the coordinates $x^\mu$ to some new coordinates $\bar x^\mu$ so that the metric $\bar g_{\mu\nu}$ has no zeros or singularities in the region of interest. Thus 
\begin{align}
\mathcal{Z}_\alpha &= \sqrt{\bar g(\alpha)} (2\pi k_BT)^2\det\left[\frac{\partial^2 E(\bar x)}{\partial \bar x^\mu\partial \bar x^\nu}\right]^{-\frac12}\exp\left(-\frac{E_\alpha}{k_BT}\right)\\
&=\left.\sqrt{\frac{\bar g(\alpha)(2\pi k_BT)^4}{\eta_1\eta_2\eta_3\eta_4}}\right|_{\bar x = \bar x(\alpha)}\exp\left(-\frac{E_\alpha}{k_BT}\right),\label{Zalpha}
\end{align}
where $\eta_i$ denotes the eigenvalues of the Hessian matrix $\left.\partial^2 E(\bar x)/\partial \bar x^\mu\partial \bar x^\nu\right|_{\bar x = \bar x(\alpha)}$.
Substituting Eq.~\eqref{Zalpha} into Eq.~\eqref{rateCoefficient} gives the result in Eq.~\eqref{attemptFrequency}
\begin{align}
r_{\alpha\beta}= \frac{\alpha_0}{1+\alpha_0^2}\sqrt{\frac{\tilde{g}(s)}{\bar g(\alpha)}} \frac{\gamma_B}m  G_U| \kappa|\sqrt{\frac{1}{2\pi k_BT}\frac{\eta_1\eta_2\eta_3\eta_4}{|\lambda_1\lambda_2\lambda_4|}} \,\exp\left(-\frac{(E_s-E_\alpha)}{k_BT}\right).
\end{align}

\end{document}